\documentclass[a4paper,fleqn,usenatbib]{mnras}

\usepackage{newtxtext,newtxmath}

\usepackage[T1]{fontenc}
\usepackage{ae,aecompl}
\usepackage{bm}

\usepackage{graphicx}	
\usepackage{amsmath}	
\usepackage{amssymb}	
\usepackage{xcolor}
\usepackage{physics}
\usepackage{mathtools}
\usepackage{placeins}

\definecolor{midblue}{HTML}{1A7282} 
\definecolor{darkblue}{HTML}{114A56}
\definecolor{purple}{HTML}{592441}
\definecolor{lightpurple}{HTML}{875F74}
\defcitealias{Aab:2018chp}{A18} 
\defcitealias{Soiaporn:2012ev}{S13} 

\title[Constraining source-UHECR associations]{Impact of using the ultra-high-energy cosmic ray arrival energies to constrain source associations}

\author[F. Capel \& D. J. Mortlock]{
Francesca Capel$^{1,2}$\thanks{E-mail: capel@kth.se (FC)}
and Daniel J. Mortlock$^{3,4,5}$
\\
$^{1}$The Oskar Klein Centre for Cosmoparticle Physics, SE-106 91 Stockholm, Sweden \\
$^{2}$Department of Physics, KTH Royal Institute of Technology, AlbaNova, SE-106 91 Stockholm, Sweden \\
$^{3}$Astrophysics Group, Imperial College London, Blackett Laboratory, Prince Consort Road, London SW7 2AZ, UK \\
$^{4}$Statistics Section, Department of Mathematics, Imperial College London, London SW7 2AZ, UK \\
$^{5}$Department of Astronomy, Stockholm University, AlbaNova, SE-106 91 Stockholm, Sweden
}

\date{Accepted XXX. Received YYY; in original form ZZZ}

\pubyear{2018}

\begin{document}

\hypersetup{
  colorlinks  = true, 
  urlcolor = purple, 
  linkcolor = darkblue, 
  citecolor = darkblue, 
}

\label{firstpage}
\pagerange{\pageref{firstpage}--\pageref{lastpage}}
\maketitle

\begin{abstract}
We present a Bayesian hierarchical model which enables a joint fit of the ultra-high-energy cosmic ray (UHECR) energy spectrum and arrival directions within the context of a physical model for the UHECR phenomenology. In this way, possible associations with astrophysical source populations can be assessed in a physically and statistically principled manner. The importance of including the UHECR energy data and detection effects is demonstrated through simulation studies, showing that the effective GZK horizon is significantly extended for typical reconstruction uncertainties. We also verify the ability of the model to fit and recover physical parameters from {\tt CRPropa~3} simulations. Finally, the model is used to assess the fraction of the the publicly available dataset of 231 UHECRs detected by the Pierre Auger Observatory (PAO) which are associated with the \emph{Fermi}-LAT 2FHL catalogue, a set of starburst galaxies and \emph{Swift}-BAT hard X-ray sources. We find association fractions of $9.5^{+2.4}_{-5.9}$,  $22.7^{+6.6}_{-12.4}$ and $22.8^{+6.6}_{-8.0}$ per cent for the 2FHL, starburst galaxies and \emph{Swift}-BAT catalogues respectively. 
\end{abstract}



\section{Introduction}
\label{sec:intro}

UHECRs are the most energetic particles ever detected with energies of above $\sim 10^{18}$~eV (1 EeV) and their origin remains an open question (see \citealt{Kotera:2011kva}, \citealt{Aloisio:2017qoo} and \citealt{Anchordoqui:2018wm} for recent reviews). UHECRs are extremely rare, with a flux of around 1 particle~km$^{-2}$~century$^{-1}$ above 50 EeV, but can be detected via air showers resulting from their interaction with the Earth's atmosphere. The development of large ground-based air shower detectors has allowed for measurements of the UHECR energy spectrum, arrival directions and mass composition. The two largest experiments, the Pierre Auger Observatory (\citealt{Collaboration:2015br}) and the Telescope Array Project (TA, \citealt{AbuZayyad:2012hi, Tokuno:2011km}), have now detected several thousands of UHECRs. 

Despite the increase in available data, the study of UHECRs remains challenging due to the complexity of the physical processes involved in their acceleration, propagation and detection. A range of extra-Galactic astrophysical sources are postulated, including non-relativistic shocks in galaxy clusters \citep{Norman:ua, Ryu:2003kj}, relativistic shocks in active galactic nuclei (AGNs, \citealt{Biermann:wt, Dermer:2009cu}), gamma-ray bursts (GRBs, \citealt{Waxman:1995jg, Vietri:1995hs}), and strong electric fields present in pulsars \citep{Blasi:2000he, Fang:2012dc}. As the UHECRs propagate from their sources to Earth, they experience considerable deflections due to the effects of Galactic and extra-Galactic magnetic fields, as well as interactions with photons of the cosmic microwave background (CMB) and the extra-Galactic background light (EBL). The energy losses through interactions with the CMB give rise to the GZK effect \citep{Greisen:1966jv, Zatsepin:1966jv}, implying that UHECRs detected at Earth with energies greater than 50~EeV must come from within a $\sim$~250 Mpc horizon. 

As UHECRs are expected to come from relatively nearby, and their high energies mean that they should experience small deflections of less than $\sim10^{\circ}$, many attempts have been made to find angular associations between UHECR arrival directions and potential sources (see e.g. \citealt{Stanev:1995ja, Tinyakov:2001de, AlvarezMuniz:2002hh, Collaboration:2007bp, Abreu:2010ce, Kim:2011ku, Oikonomou:2013hy, Aab:2015js} and references therein). Most previous work has taken a hypothesis testing approach based on purely spatial correlations. A general consensus has not been reached, with differing conclusions on possible associations based on different UHECR datasets, source catalogues and statistical methods. Most recently, the PAO Collaboration reported a correlation between the arrival directions of UHECRs with arrival energies greater than 39~EeV and the positions of 23 starburst galaxies within 250~Mpc (\citealt{Aab:2018chp}, hereafter \citetalias{Aab:2018chp}). The TA Collaboration followed up on this result by repeating the analysis on their UHECR dataset, finding a result that is both consistent with isotropy and the starburst galaxy catalogue \citep{Collaboration:2018ws}. 

While studying the UHECR arrival directions in isolation can provide some insight, it is crucial to consider the complementary information provided by the UHECR arrival energies in order to draw physical conclusions. We do not expect the highest energy UHECRs to come from distant sources or to be highly deflected. Within the context of a physical model for UHECR acceleration, propagation and detection, the UHECR energies provide an additional constraint that avoids unphysical source-UHECR angular associations and allows us to assess whether a population of sources can realistically represent the observed dataset. Including the energy data also allows us to directly constrain the underlying physical processes, such as the injection spectrum of the source population and the magnetic fields responsible for the energy-dependent deflections of UHECRs.  

Building a comprehensive statistical model for the UHECR observations is non-trivial due to the high numbers of parameters and uncertainties involved, but can be achieved within the framework of Bayesian hierarchical modelling. Previous work has made headway in this direction, \citet{Watson:2010xx} and \citet{Khanin:2016er} developed a two-component parametric model characterised by source and background rates, modelling UHECR arrival directions as drawn from a binned Poisson intensity on the celestial sphere. Their model also accounts for UHECR energy losses during propagation in terms of weighting the sources, but does not include the UHECR energies explicitly in the likelihood. In \citet{Khanin:2016er}, the model is also applied to the public PAO dataset of 69 UHECR events to a variety of catalogues, finding the largest association fractions of $0.25^{+0.09}_{-0.08}$ and $0.24^{+0.12}_{-0.10}$ for the \emph{Swift}-BAT hard X-ray catalogue and the 2MASS Redshift Survey respectively. Further, \citet{Soiaporn:2012ev} (hereafter \citetalias{Soiaporn:2012ev}) present an extendable hierarchical framework in which they derive an expression for the posterior distribution by modelling UHECR arrival directions as an inhomogeneous Poisson process on the celestial sphere. Their model was applied to the public dataset of 69 UHECR events detected by PAO and a volume-limited sample of 17 nearby AGNs \citep{Goulding:2010pq}, finding small but non-zero association fractions of up to $\sim$~0.2. 

Motivated by the desire to include the extra information provided by the UHECR energies, we present a joint model for the UHECR energies and arrival directions. This builds on the formalism of \citetalias{Soiaporn:2012ev} and includes the ideas of \citet{Watson:2010xx} and \citet{Khanin:2016er}. We derive the model in Sections \ref{sec:physics} and \ref{sec:model}, then validate the model through simulations in Section \ref{sec:sim}. Following this, we apply the model to the most recent public PAO dataset. A range of source catalogues are tested against this dataset and the results are presented in Section \ref{sec:application}. The code used in this work is available online\footnote{\url{https://github.com/cescalara/uhecr_model} (to be made available upon publication)}.


\section{UHECR model}
\label{sec:physics}

\begin{figure}
 \includegraphics[width=\columnwidth]{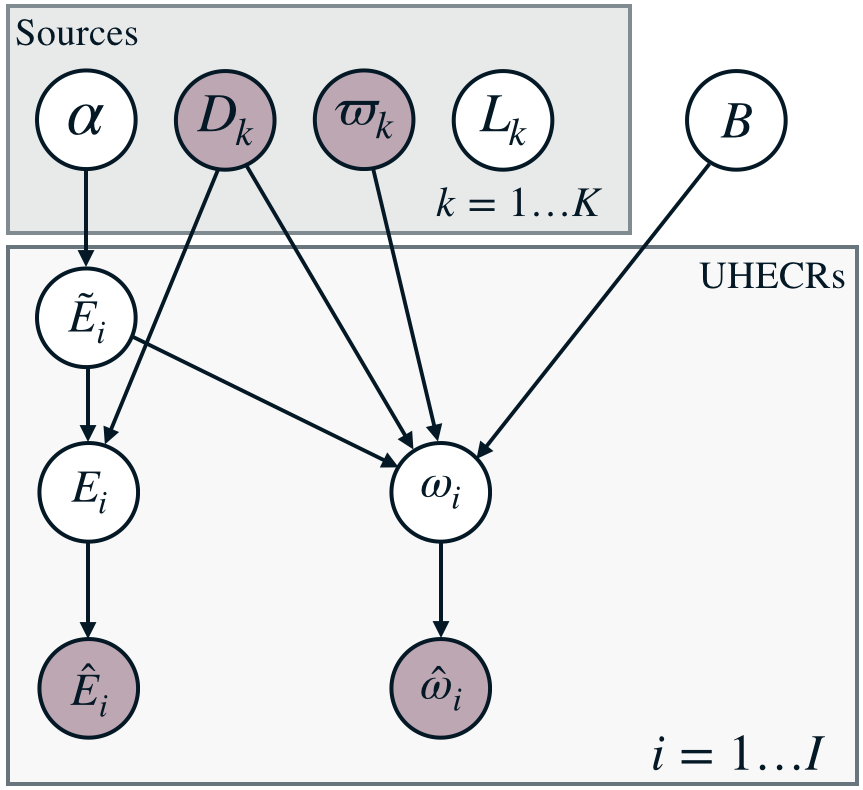}
 \caption{Directed acyclic graph showing the hierarchical model for UHECR energies and arrival directions. Data and parameters are shown using shaded and unshaded circles respectively. The two boxes gather together the source and UHECR populations. Parameters outside of the boxes are global and are not associated with either population. The notation is explained in the text.}
 \label{fig:dag}
\end{figure}

For a given set of detected UHECRs with measured energies and arrival directions, a catalogue of potential sources and some knowledge on the physics of UHECR acceleration and propagation, the statistical problem is to quantify the probability of association between the UHECRs and the sources. This can be done by breaking the problem down into different levels and building a hierarchy based on the underlying physical processes. We consider the case of $K$ sources, indexed by $k$, and $I$ UHECRs, indexed by $i$, as shown in Figure \ref{fig:dag}. Following \citetalias{Soiaporn:2012ev}, the model has 4 levels: the source population (Section~\ref{sec:physics:sources}), and the UHECR acceleration, propagation and detection (Sections~\ref{sec:physics:acc}-\ref{sec:physics:detection}). We also highlight some key implications of the model in more detail in Section~\ref{sec:physics:impact_det}

\subsection{Source population}
\label{sec:physics:sources}

We assume a population of steady-state sources, such that the UHECR flux at Earth does not change over the time-scales of the available observations. In this way, the production rate from source $k$ can be modelled as a homogeneous Poisson point process in time, characterised by its rate, $L_k$ in units of time$^{-1}$. The $k^{\rm{th}}$ source is defined by a direction on the celestial sphere, $\varpi_k$, and a distance from the Earth, $D_k$. We consider equal luminosities for all sources, such that $F_k = L / (4 \pi D_k^2)$, for $k = 1$ to $K$. 

We also include an isotropic background component to represent UHECRs from a population of distant, unresolved sources, or sources not present in the chosen catalogue. This component is labelled by $k = 0$ and is described by $F_0$, the total flux due to this background population in units of area$^{-1}$~time$^{-1}$. 


\subsection{Acceleration}
\label{sec:physics:acc}

The complex acceleration physics that produces UHECRs is summarised by modelling the emission spectrum of UHECRs with a power law. The number of UHECRs at a certain energy is described by ${\mathrm{d}N / \mathrm{d}E \propto E^{-\alpha}}$, where the spectral index $\alpha$ is expected to be $\sim$~2 from Fermi acceleration arguments \citep{Bell:1978is, Baring:2001gf}. The spectrum is normalised such that the emission rate per unit energy of UHECRs from source $k$ is given by
\begin{equation}
\frac{\dd{N_k}}{\dd{E}\dd{t}} = L \frac{\alpha - 1}{E_{\rm{min}}} \left( \frac{E}{E_{\rm{min}}} \right)^{-\alpha},	
\label{eqn:Enorm}
\end{equation}
where $E_{\rm{min}}$ is the minimum UHECR emission energy and $L$ is the rate at which source $k$ emits UHECRs with $E > E_{\rm{min}}$. We use $E_{\rm{min}} = 52$~EeV when considering the PAO dataset in Section~\ref{sec:application} and fits to the simulated data in Section~\ref{sec:sim} as it is the minimum energy of the sample data. However, when generating simulated datasets in Section~\ref{sec:sim}, we use $E_{\rm{min}} = 20$~EeV in order to take into account the effect of the uncertainty of the energy reconstruction, as described in Section~\ref{sec:physics:detection}.


\subsection{Composition}
\label{sec:physics:composition}

The UHECR composition is challenging to reconstruct from air shower measurements, with the depth of the shower maximum, $X_{\rm{max}}$ and its fluctuations $\sigma(X_{\rm{max}})$ most often used to measure the hadronic component of the shower. The most recent observations are inconclusive, with TA favouring a lighter, proton composition \citep{abbasi:dh} and the PAO a mixed composition \citep{Aab:2016ci} at the highest energies. However, these results are actually consistent within the uncertainties present \citep{Hanlon:2018ua}. 

We assume that all UHECRs are protons, allowing us to implement a relatively simple model for the UHECR propagation to demonstrate our framework. The model presented here could be extended to include a mixed UHECR composition as described in Section~\ref{sec:conclusion}.

  
\subsection{Propagation}
\label{sec:physics:propa}

During their propagation, UHECRs lose energy (Section~\ref{sec:physics:Eloss}) and are deflected by magnetic fields (Section~\ref{sec:physics:Bfield}).

\subsubsection{Energy losses}
\label{sec:physics:Eloss}

\begin{figure}
\includegraphics[width=\columnwidth]{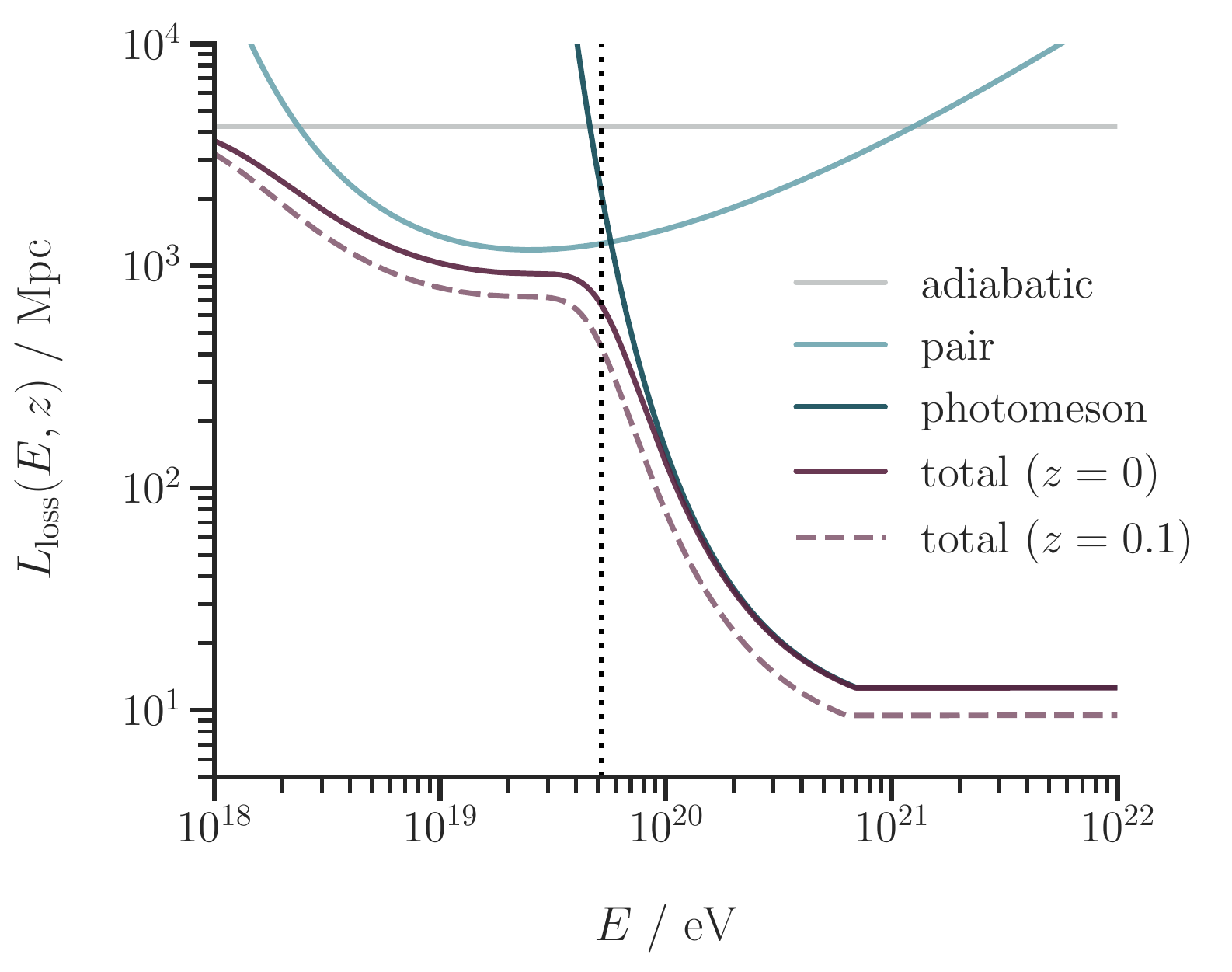}
\caption{The total $L_{\mathrm{loss}}$ is shown as a function of energy for the cases of redshift $z = 0$ and $z=0.1$, corresponding to $\sim$~400~Mpc and thus covering the GZK horizon. The 3 different processes are also shown separately for $z=0$ (c.f. figure~2 in \citealt{DeDomenico:2012ev}). The dotted line shows $E = 52 \times 10^{18}$~eV, the minimum energy of the PAO UHECR dataset used in this work.}
\label{fig:loss_length}	
\end{figure}

We make use of the continuous energy loss approximation \citep{Berezinskii:uj, Chodorowski:1992ei, Anchordoqui:1996ru, DeDomenico:2012ev} for energy losses during propagation. Under this approximation, the arrival energy of UHECRs can be calculated by solving
\begin{equation}
\frac{ \dd{E} }{ \dd{z} } = - \frac{E}{L_{\rm{loss}}(E, z)},
\label{eqn:E_loss}
\end{equation}
where $z$ is the redshift of the UHECRs and $L_{\rm{loss}}(E, z)$ is the loss length. There are three distinct processes which contribute significantly to the energy losses of protons for the energies considered here. Therefore $L_{\rm{loss}}(E, z)$ can be expressed as
\begin{equation}
L^{-1}_{\rm{loss}}(E, z) = \frac{ \dd{t} }{ \dd{z} } [ \beta_{\pi}(E, z) + \beta_{e^{\pm}}(E, z) + \beta_{\mathrm{adi}}(z) ],
\end{equation}
where the $\beta$ terms are the losses per unit time corresponding to the following processes:
\begin{enumerate}
\item photomeson production (e.g. $p + \gamma \rightarrow n + \pi^+$) due to high-energy protons interacting with the CMB (the GZK effect);
\item Bethe-Heitler pair production ($p + \gamma \rightarrow p + e^+ + e^-$), also due to CMB interactions;
\item the adiabatic expansion of the Universe.
\end{enumerate}
Using the parametrization described in \citet{DeDomenico:2012ev}, $L_{\rm{loss}}$ is as shown as a function of energy in Figure \ref{fig:loss_length}. For the typical energies considered here ($E > 52$~EeV) the dominant process is photomeson production. While continuous energy losses are a reasonable approximation for protons in this case, they do not account for spread in the resulting energy spectrum due to the kinematics of the interaction and the Poisson noise in the number of interactions \citep{Hill:1983mk, Achterberg:1999vr}. 

For the energy loss calculations used throughout this work, we assume ${z = D H_0/ c}$, where $D$ is the luminosity distance in Mpc. This approximation is valid for the small redshifts ${(z < 0.06)}$ considered. For all calculations involving cosmological parameters we assume ${H_0 = 70 \ \mathrm{km} \ \mathrm{s}^{-1} \ \mathrm{Mpc}^{-1}}$, and a standard $\Lambda$CDM model with $\Omega_\mathrm{m} = 0.3$ and $\Omega_\Lambda = 0.7$. 

\subsubsection{Magnetic deflections}
\label{sec:physics:Bfield}

UHECRs undergo deflections due to the various magnetic fields encountered during their propagation. The magnitude of these deflections is difficult to quantify, given the lack of knowledge about the strength and structure of the Galactic and extra-Galactic magnetic fields (see e.g \citealt{Vallee:2004ep} and \citealt{Kulsrud:2008du} for an overview). Progress has been made in modelling the Galactic magnetic field \citep{Jansson:2012epa}. The regular component of the Galactic magnetic field can act as a lens, leading to correlated deflections of UHECRs which are a function of the position on the sky \citep{Waxman:1996zn}. Broadly speaking, the local Galactic magnetic field can be described with a regular component of roughly 3 $\upmu$G and a Kolmogorov-type turbulent component of similar magnitude on length scales of $\sim 100$~kpc  \citep{Durrer:2013ec}. Extra-Galactic magnetic fields are less constrained by observations and can vary over many orders of magnitude in both strength and coherence length, from smaller than $\sim$~nG in voids over 1~Mpc scales to up to $\sim$~$\upmu$G in 100~kpc scale galaxy clusters \citep{Kotera:2008gxa, Aharonian:2010cr, Durrer:2013ec}.

Following \citet{Harari:2002dy}, we model the total magnetic field as a random Gaussian field with zero mean and root mean square field strength $\bar{B}$ and do not separate the Galactic and extra-Galactic contributions. In the limit of small deflections and the distance travelled being much larger than the coherence length (${D \gg l_{\rm{c}}}$) the deflections of UHECRs follow a probability distribution characterised by zero mean and root mean square angular scale, given by
\begin{equation}
\bar{\theta} \approx 2.3^{\circ} \left( \frac{E}{50 \ \rm{EeV}} \right)^{-1} \left( \frac{\bar{B}}{1 \ \rm{nG}} \right) \left( \frac{D}{10 \ \rm{Mpc}} \right)^{1/2} \left( \frac{l_{\rm{c}}}{1 \ \rm{Mpc}} \right)^{1/2},
\label{eqn:deflection}
\end{equation}
where the quantities have been scaled to typical extra-Galactic values and a pure proton composition is assumed. Given the uncertainties in modelling the magnetic fields, we leave $\bar{B}$ as a free parameter in the model, but choose a fixed ${l_{\rm{c}} = 1}$~Mpc to avoid degeneracy and represent extra-Galactic scales.


\subsection{Detection}
\label{sec:physics:detection}

The exposure of a ground-based UHECR observatory such as PAO or TA can be calculated by considering the geometry of the system. For an observatory at latitude $a_0$ that can detect UHECRs arriving with zenith angles up to $\theta_{\rm{m}}$, the exposure as a function of position on the sky, $\epsilon(\omega)$ is given by
\begin{equation}
\epsilon(\omega) = \frac{ \alpha_{\rm{T}} }{ M } m(\omega),
\end{equation}
where $\omega$ is a unit vector describing the UHECR arrival direction and $\alpha_{\rm{T}}$ is the total exposure of the observatory in units of area~$\times$~time~$\times$~solid~angle and $M = \int \dd{\omega} m(\omega)$. The arrival direction, $\omega$ can also be expressed in terms of right ascension and declination. The time-averaged exposure over 1 sidereal day is purely a function of declination, $\delta$, and is given by
\begin{equation}
m(\omega) = m(\delta) = \cos(a_0)\cos(\delta)\sin(\alpha_{\rm{m}}) + \alpha_{\rm{m}} \sin(a_0) \sin(\delta),
\end{equation}
where $\alpha_{\rm{m}}$ and the $\theta_{\rm{m}}$ dependence are as described in \citet{Sommers:2000bo}. $\epsilon(\omega)$ is proportional to the probability that a UHECR is detected given that it has arrival direction $\omega$. For the PAO, we have $a_0 = 35.2^{\circ}$, $\theta_{\rm{m}} = 80^{\circ}$ and $\alpha_{\rm{T}} = 66,000$~km$^{2}$~sr~yr, as described in \citet{Aab:2015js}.

In addition to the exposure effects, the reconstruction uncertainties in the measured UHECR energies and arrival directions are also important. The uncertainty in the energy reconstruction is reported as a fraction of the arrival energy of the UHECR, $\sigma_E = f_E E$ where $f_E$ is between 0 and 1 \citep{Collaboration:2015br}. We model the detected energies as being Gaussian distributed 
\begin{equation}
P(\hat{E} | E, \sigma_E) \sim \mathcal{N}(E, f_E^2 E^2),
\label{eqn:E_uncert}
\end{equation}
where $\hat{E}$ is the detected energy and $E$ is the true arrival energy. Similarly, the uncertainty in the reconstruction of the arrival direction is modelled using a von Mises-Fisher (vMF, \citealt{Fisher:1953gp}) distribution centred on the true arrival direction and with the spread characterised by the dimensionless parameter $\kappa_{\rm{d}}$ as
\begin{equation}
P(\hat{\omega} | \omega, \kappa_{\rm{d}}) = \frac{\kappa_{\rm{d}}}{4 \pi \sinh{(\kappa_{\rm{d}}})} e^{\kappa_{\rm{d}} \hat{\omega} \cdot \omega},
\end{equation}
where $\hat{\omega}$ is the detected arrival direction, $\omega$ is the true arrival direction and $\kappa_{\rm{d}}$ is inversely related to the width of the distribution. If ${\kappa_{\rm{d}} \rightarrow 0}$, the vMF distribution becomes uniform on the sphere; if ${\kappa_{\rm{d}} \gg 1}$, the vMF distribution approaches a bivariate Gaussian distribution near the mode. As the angular reconstruction uncertainty, $\sigma_\omega$, is often reported in units of degrees, we can convert this to $\kappa_{\rm{d}}$ in analogy with the `1~$\sigma$' region of a Gaussian distribution. As highlighted in \citetalias{Soiaporn:2012ev}, we have
\begin{equation}
\kappa_{\rm{d}} \approx 7552 \left( \frac{\sigma_{\omega}}{1^{\circ}} \right)^{-2}.
\label{eqn:convMF}
\end{equation}
The reconstruction uncertainties of the PAO are reported to be $f_E \leq$~0.12 for the energies \citep{Collaboration:2015br, Aab:2015js} and ${\sigma_\omega \leq 0.9^{\circ}}$ for the arrival directions \citep{Bonifazi:2009hm}. As we do not have heteroscedastic event-by-event uncertainties, these maximum values are adopted in our analysis. We do not consider the additional systematic uncertainty of $\sim$~14~per~cent on the absolute energy scale.


\subsection{Impact of detection uncertainties}
\label{sec:physics:impact_det}

The number of UHECRs detected per unit area per unit time from a particular source, $N_k$, is proportional to the rate of UHECR emitted by that source, $L$, but must also take into account the source energy spectrum and the effects of UHECR energy losses. The energy loss model described in Section \ref{sec:physics:Eloss} can be used to calculate the corresponding energy of a UHECR at the source, given its arrival energy. In this way, we can define a threshold source energy, $\tilde{E}_{\rm{th}}$ that corresponds to a threshold arrival energy at Earth, $E_{\rm{th}}$ for a source at a given distance, $D_k$. Along with Equation \ref{eqn:Enorm}, this allows us to define the number of UHECR per unit time and area as
\begin{equation}
\frac{ \dd{N_k (E > E_{\rm{th}} }) }{ \dd{t} \dd{A} } = \frac{L}{4 \pi D_k^2} \left( \frac{\tilde{E}_{\rm{th}}^{k}}{E_{\rm{min}}} \right)^{1 - \alpha}.
\label{eqn:flux_gzk}
\end{equation}

In order to apply Equation \ref{eqn:flux_gzk} to a sample that has been selected according to the detected UHECR energies, as will be necessary when working with the available UHECR data, we must assume that a hard threshold on the detected energies corresponds to a hard threshold on the source energies. This is not true, due to the energy detection uncertainties (Equation \ref{eqn:E_uncert}). We approximate these effects by truncating the Gaussian distribution in Equation \ref{eqn:E_uncert} at $E_{\rm{th}}$, which allows for the use of Equation \ref{eqn:flux_gzk}. The validity of the approximation is determined by the ability of the model to fit data simulated with realistic detection effects, as discussed in Section \ref{sec:sim}. Elsewhere in the model, we do not `invert' the detected UHECR energies in order to infer their energies at the source, but introduce latent initial energy parameters, $\tilde{E}_i$, to fit for the UHECR source energies, including the detection effects.
	
\begin{figure}
\includegraphics[width=\columnwidth]{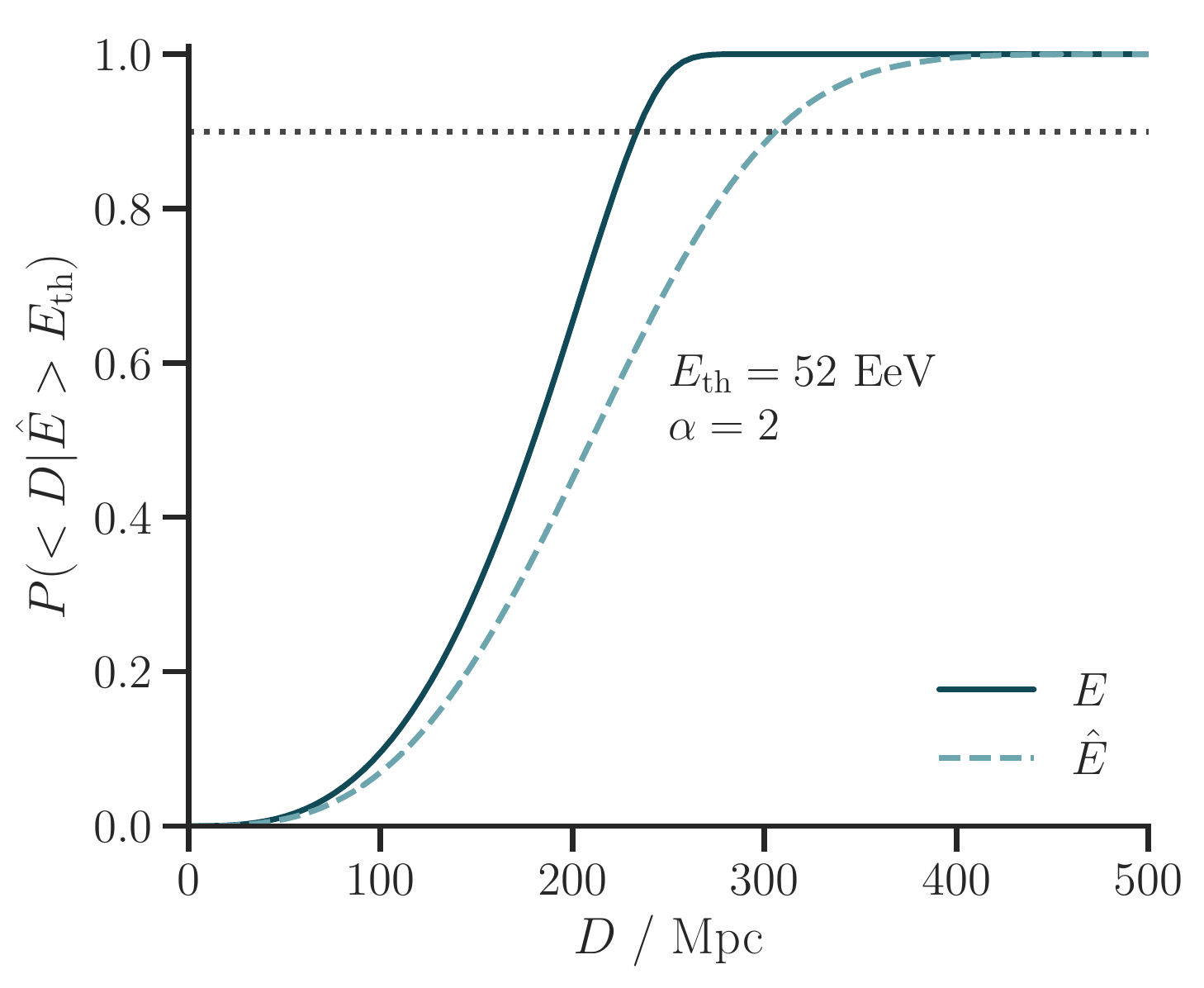}
 \caption{The probability that a UHECR is detected from within a distance $D$, given that it is detected with an energy $\hat{E} > E_{\rm{th}}$. The solid and dashed lines show the results with and without detection uncertainties respectively for the case of $E_{\rm{th}} = 52$ and $\alpha = 2$. The grey dotted line shows ${P(< D | \hat{E} > E_{\rm{th}})=0.9}$, a value which can be used to define the GZK horizon \citep{Harari:2006uy}.}
 \label{fig:Edet_effect}
\end{figure}

The uncertainty on the arrival energies means that some events in a UHECR sample will be lower energy events which are reconstructed with energies above the threshold, resulting in an effective extension of the GZK horizon. To quantify the impact of this, we consider the probability that a UHECR comes from within a distance $D$, given that it is detected with $\hat{E} > E_{\rm{th}}$. This can be expressed as
\begin{equation}
\begin{split}
P(< D | \hat{E} > E_{\rm{th}}) & = \int_0^D \dd{D'} P(D' | \hat{E} > E_{\rm{th}}) \\
& = \int_0^D \dd{D'} P(D') P(\hat{E} > E_{\rm{th}} | D'), 
\end{split}
\label{eqn:P_D_E}
\end{equation}
where $P(\hat{E} > E_{\rm{th}} | D)$ can be estimated through simulations and $P(D) \propto D^2$ for the case of homogeneously distributed sources in the nearby universe. We evaluate $P(< D | \hat{E} > E_{\rm{th}})$, both with and without detection uncertainties and the results are shown in Figure \ref{fig:Edet_effect}. For an uncertainty of 12 per cent on the reconstructed energies and $E_{\rm{th}} = 52$~EeV, we expect around 30 per cent of UHECR in the sample to have come from beyond 250~Mpc. Therefore, if one only considers sources within 250~Mpc, we expect to find a background contribution of 30~per~cent to the observed UHECR flux due to the contribution of sources beyond the GZK horizon, which would be negligible if assuming that we observe $E$ instead of $\hat{E}$. 

\begin{figure}
\includegraphics[width=\columnwidth]{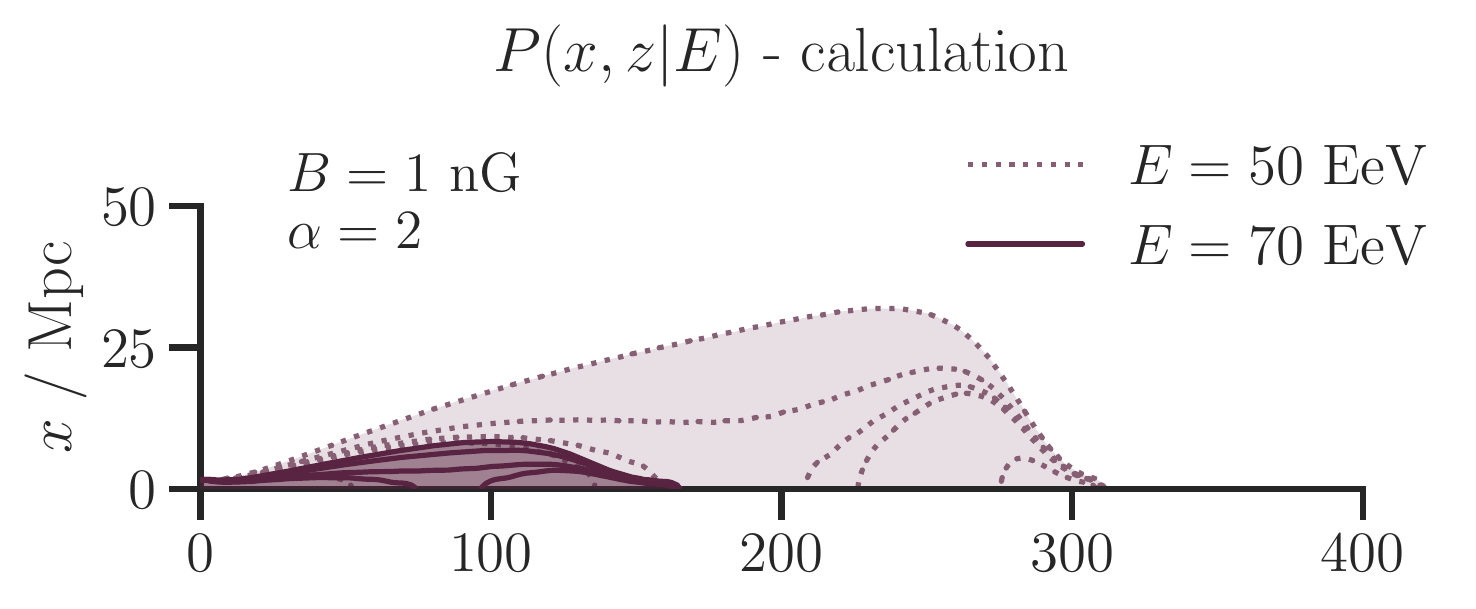}
\includegraphics[width=\columnwidth]{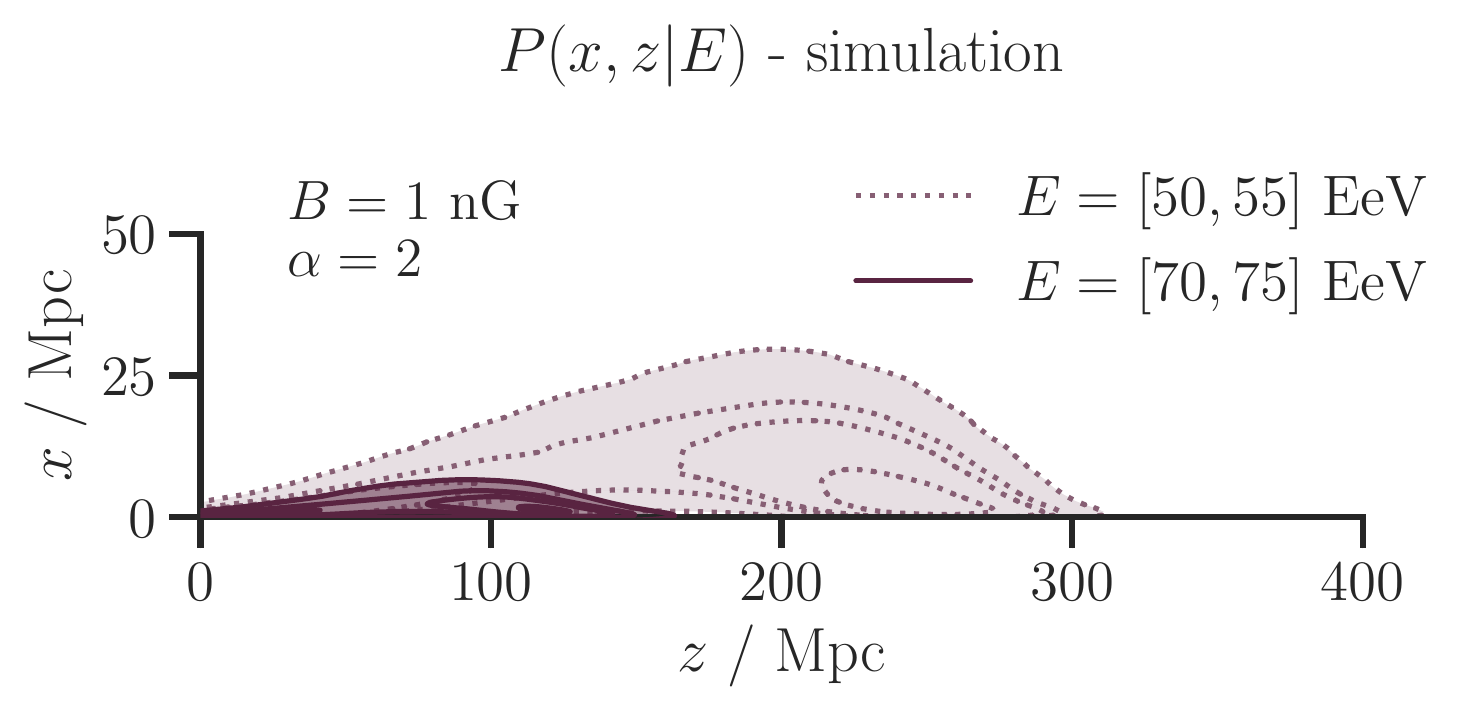}
\caption{The joint probability that a UHECR originates from $(x, z)$ given that it is detected with arrival energy $E$, with rotational symmetry around the $z$-axis. The top panel shows the result for the expression in Equation~\ref{eqn:P_x_z} for the case of $\bar{B} = 1$~nG and $\alpha = 2$. The lower panel shows the same probability evaluated by using simulations and selecting arrival energies in the ranges of [50, 55] and [70,75]~EeV. }
\label{fig:P_x_z_calc}
\end{figure}

To summarise the implications of the model for potential UHECR sources, we also consider the joint probability of the UHECR deflections and the source distances, conditioned on different detected energies. To make this easier to visualise, we express the result in terms of Cartesian coordinates with the $z$-axis representing the arrival direction of a UHECR and the $xy$-plane the surface onto which it arrives. Using the same assumptions on the source population as mentioned above, this can be expressed as
\begin{equation}
\begin{split}
P(x, y = 0, z | \hat{E}, \hat{\omega} = 0, S) & = \frac{ P(D, \theta, \phi, \tilde{E} | \hat{E}, \hat{\omega} = 0, S) }{D^2 \sin(\theta)} \\
& = \frac{ P(D, \theta, | \hat{E}, \hat{\omega} = 0, S)  }{D^2 \sin(\theta)}, 
\end{split}
\end{equation}
where $S$ signifies that the UHECR is detected at Earth, $\hat{\omega}$ are the detected UHECR arrival directions, $\theta$ are the deflections of the detected UHECR from the source location and $D$ and $\phi$ make up the spherical coordinate system. In the last step we have marginalised over $\phi$ and $\tilde{E}$. If we ignore the detection uncertainties, thus conditioning on $(E, \omega)$ instead of $(\hat{E}, \hat{\omega})$, this can be expressed analytically as
\begin{multline}
P(x, y = 0, z | E, \omega = 0, S) \propto \\ 
D^{-3} \tilde{E}(E, D)^{2 - \alpha} \frac{e^{\theta_0^{-2} \tilde{E}(E, D)^2 D^{-1} \cos(\theta)}}{ \sinh[\theta_0^{-2} \tilde{E}(E, D)^2] } \left| \frac{ \dd{E(\tilde{E}', D)} }{ \dd{\tilde{E}'} } \right|_{\tilde{E}' = \tilde{E}(E, D)},
\label{eqn:P_x_z}
\end{multline}
where $\theta_0$ is the constant part of the $\bar{\theta}$ dependence given in Equation~\ref{eqn:deflection} and we have used the vMF distribution to represent the UHECR deflections with $\bar{B} = 1$~nG. We plot this distribution in the top panel of Figure~\ref{fig:P_x_z_calc} for the case of $E = 50$ and $E = 70$~EeV. We see that higher energy UHECRs must originate from closer sources in addition to having smaller deflections. In this way, the model reflects that there is a particle-specific GZK horizon for each UHECR that is smaller for higher energy particles. The contours show two peaks in the distribution, indicating that UHECRs are most likely to come from either $D = 0$ or towards the location of the effective GZK horizon for the energy considered. This can be understood by the fact that due to the energy losses under the continuous loss approximation, protons produced at the horizon will `pile up' at the arrival energy, whereas for nearby sources, the source spectrum is equal to the arrival spectrum and so follows the power-law form. Additionally, we expect a larger contribution from near the effective horizon. This effect is less prominent at higher energies ($E \gg E_{\rm{GZK}}$), as the energy losses become effectively exponential and the Jacobian term in Equation~\ref{eqn:P_x_z} tends to unity.

\begin{figure}
\includegraphics[width=\columnwidth]{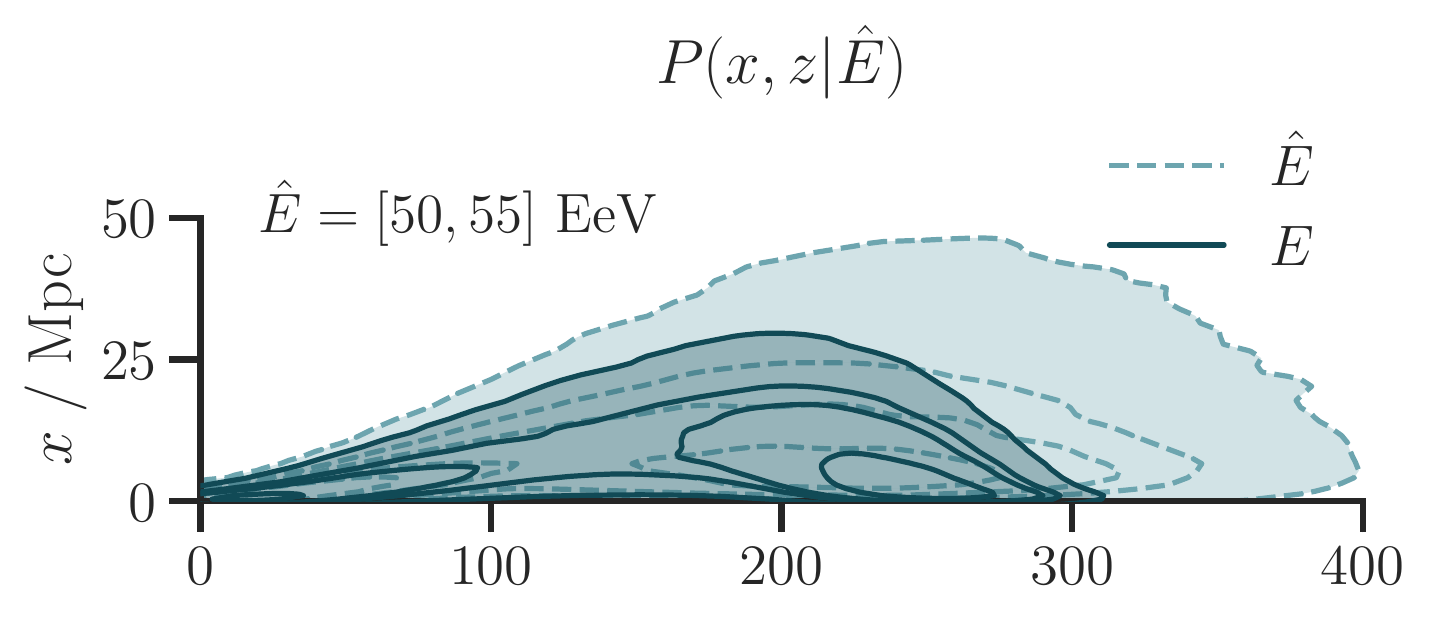}
\includegraphics[width=\columnwidth]{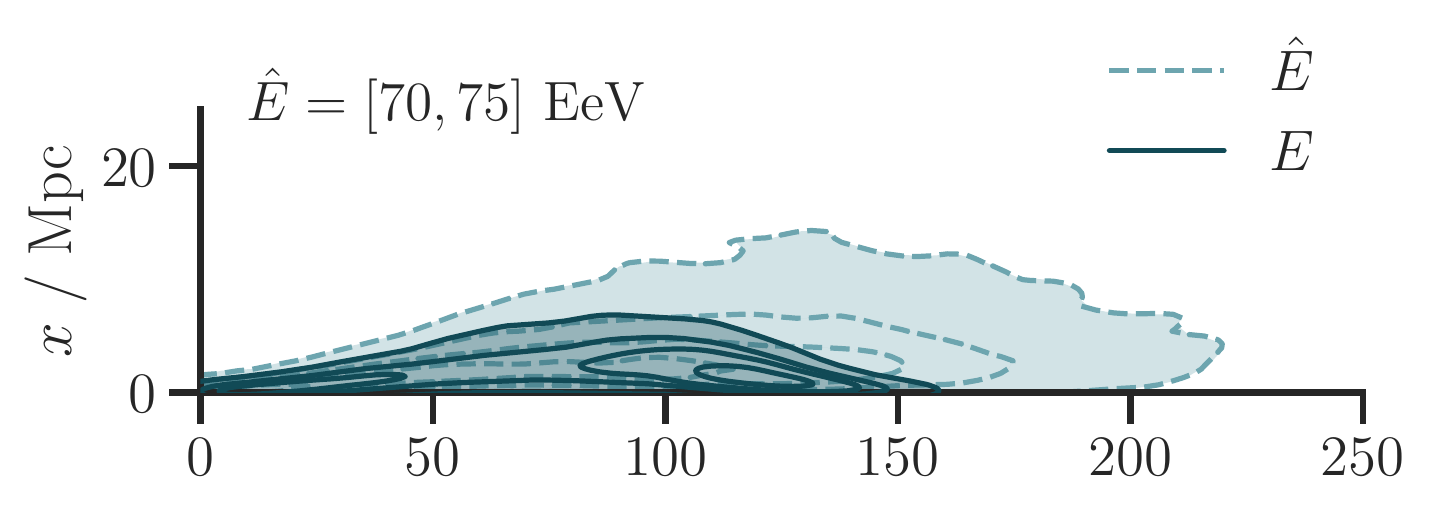}
\includegraphics[width=\columnwidth]{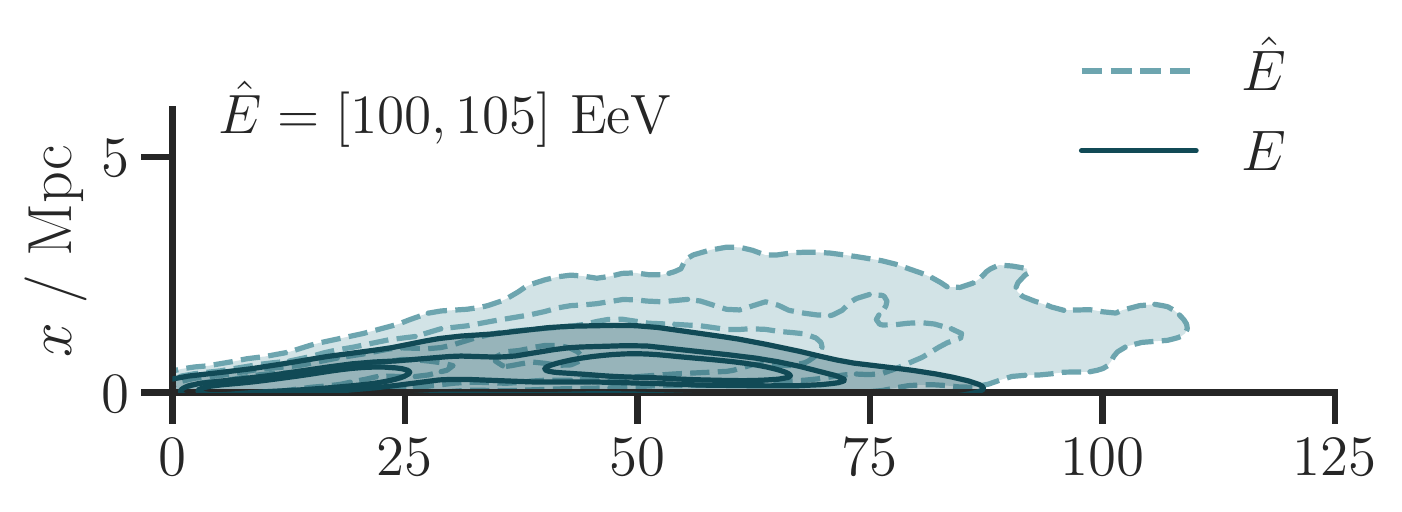}
\includegraphics[width=\columnwidth]{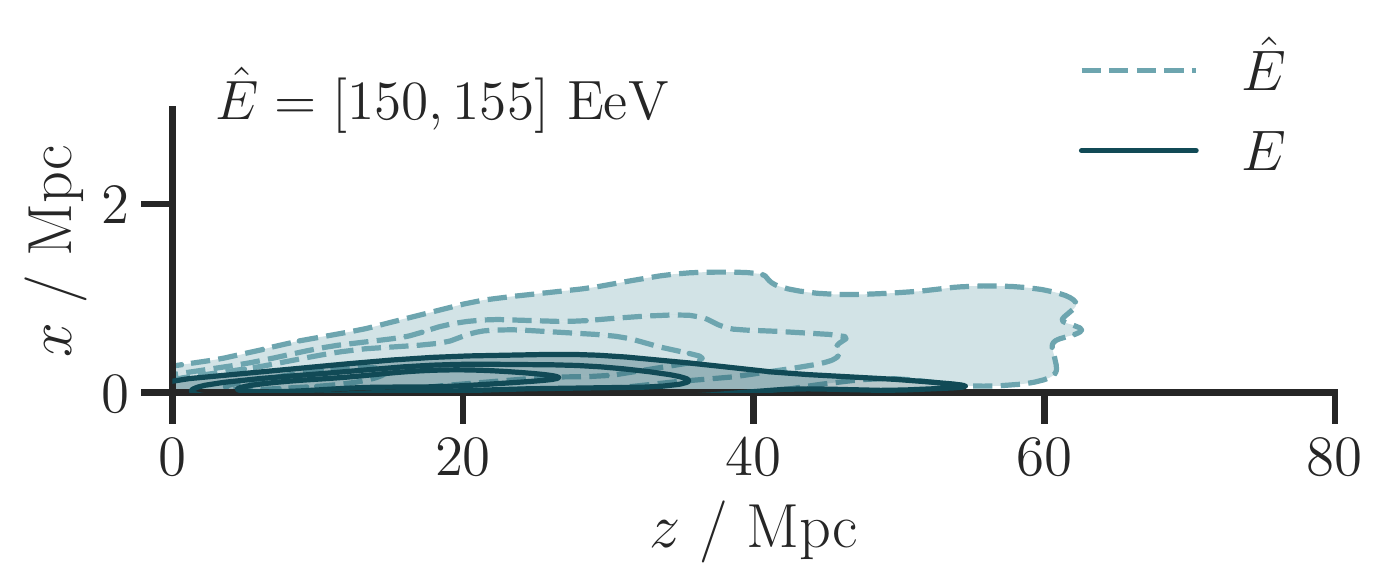}
\caption{The joint probability that a UHECR originates from $(x, z)$ given that it is detected with energy $E$ or $\hat{E}$, evaluated using simulations. The distribution has rotational symmetry around the $z$-axis. As in Figure~\ref{fig:Edet_effect}, the dashed and solid lines show the results with and without detection uncertainties respectively for the case of $\bar{B} = 1$~nG and $\alpha = 2$. The different panels from top to bottom show the case for increasing detected energy. At higher energies, the double peak visible in the distribution disappears, as discussed in the text. Note that both the $x$ and $z$ axes ranges vary in order to show the structure of the distributions.}
\label{fig:P_x_z}
\end{figure}

We evaluate $P(x, y = 0, z | \hat{E}, \hat{\omega} = 0, S)$ using simulations both with and without detection uncertainties on the UHECR energies and arrival directions. For the case without detection uncertainties, we compare with result with the calculation described above in Figure~\ref{fig:P_x_z_calc} and see that we find a good match between the two, with some small differences to be expected due to the selection of a range of arrival energies in the simulation. We then show the comparison with and without detection uncertainties in Figure~\ref{fig:P_x_z}. These distributions represent a slice through the 3D `bubble' of possible UHECR origins given that it is detected with energy $E/\hat{E}$ and direction $\omega/\hat{\omega}$ (aligned with the $z$-axis). Again, we note the importance of including the detection effects, with the distribution of possible UHECR origins extending further in both $x$ and $z$ for detection uncertainties of $\sigma_\omega = 0.9^{\circ}$ and $f_E = 0.12$, which correspond to the case of the PAO dataset used in this work. Increasing the horizon distance from, for example, 250 to 360~Mpc more than doubles the volume and hence the number of potential sources that should be considered.

\section{Statistical formalism}
\label{sec:model}

Using the above UHECR model, we derive an expression for the likelihood, which gives the probability of the observed data given a set of model parameters (Section~\ref{sec:model:likelihood}). For clarity, we highlight the main points here; the full derivation can be found in Appendix \ref{app:likelihood}. We then show how the model parameters can be used to obtain an expression for the associated fraction of UHECRs in Section~\ref{sec:model:f}. We describe the prior distributions in Section~\ref{sec:model:priors} before bringing this together with the likelihood in Section~\ref{sec:model:inference} to obtain an expression for the posterior distribution which can be evaluated numerically. Finally, we also derive a simpler model for only the UHECR arrival directions in Section~\ref{sec:model:arrdir} for comparison. 


\subsection{Likelihood}
\label{sec:model:likelihood}

Whilst the production of UHECRs can be modelled as constant in time, their detection cannot, due to the rotation of the Earth and the corresponding time dependence of the instantaneous exposure of an observatory. This gives a detection rate which is a function of both UHECR arrival directions and time, $r(\omega, t)$. The resulting likelihood function is that of an inhomogeneous Poisson point process \citep{Loredo:2004hl, PPP} and has the form
\begin{equation}
P(\hat{E}, \hat{\omega} | L, \alpha, \tilde{E}, \bar{B}) \propto e^{-\bar{N}}\prod_{i=0}^{I} r(\omega_i, t_i), 
\label{eqn:inhomopp}
\end{equation}
where $\bar{N} = \int \int \dd{\omega} \dd{t} r(\omega, t)$ is the total expected number of UHECRs in the observation period. The likelihood which we derive here has the same overall structure, but with modifications due to the fact that we model the UHECRs as being produced with labels, $\lambda_i$, which identify their sources, and initial energies, $\tilde{E}_i$. This introduces associated distributions for the labels and energies such that UHECR production can be described by a marked Poisson point process. 

In order to write down an expression for the Poisson rate, it is necessary to consider contributions to the likelihood from all possible source-UHECR combinations. This can be achieved by marginalising over the latent UHECR labels, resulting in what is effectively a mixture model over different source contributions, including the background component (see Section~\ref{sec:physics:sources}). Building on Equation \ref{eqn:inhomopp}, we have
\begin{flalign}
\label{eqn:likelihood}
& P(\hat{E}, \hat{\omega} | L, \alpha, \tilde{E}, \bar{B}) = && \\
&&&& \mathmakebox[0pt][r]{%
e^{-\bar{N}} \prod_{i = 0}^I \sum_{k = 0}^{K} \Bigg[ F_k P(\hat{\omega}_i | \varpi_k, D_k, \tilde{E}_i, \bar{B}) P(\hat{E}_i | E_i) P(E_i | \tilde{E_i}, D_k) P(\tilde{E}_i | \alpha) \Bigg].
}
\notag
\end{flalign}
Here, the $F_k$ act as weights, with $P(\hat{\omega}_i | \varpi_k, D_k, \tilde{E}_i, \bar{B})$ quantifying the probability of observing a UHECR with arrival direction $\hat{\omega}_i$, given that it comes from source $k$, with initial energy $\tilde{E}_i$ and travels through a magnetic field of mean strength $\bar{B}$. The latent $\omega_i$ have been marginalised over. The other terms in the likelihood, $P(\hat{E}_i | E_i)$, $P(E_i | \tilde{E_i}, D_k)$ and $P(\tilde{E}_i | \alpha)$ give the probability of observing a UHECR with energy $\hat{E}_i$ given that the production spectrum is a power law with spectral index $\alpha$ and that the energy losses during propagation are as described in Section \ref{sec:physics:Eloss}. 

The background component with $k = 0$ assumes no energy losses (equivalent to $D_0 = 0$~Mpc). This means that if a low associated fraction is found, the model essentially fits an isotropic distribution of arrival directions and $\alpha$ represents the arrival energy spectrum and not that of the sources. 

To calculate the expected number of events, $\bar{N}$, we must sum over the contribution of all possible sources, whilst taking into account the effect of the GZK horizon and the exposure of the detector. Energy losses are considered in the derivation of Equation \ref{eqn:flux_gzk}, which defines an expression for the flux of UHECRs from each source. We continue by defining an effective exposure factor for each source, $\epsilon_k$, which is the convolution of the source vMF distribution and the observatory exposure. This depends on both the direction of the source on the celestial sphere and the spread of the vMF distribution associated with that source. The effective exposure is hence indirectly dependent on the source UHECR energy distribution, the source distance and the magnetic field strength. This gives
\begin{equation}
\bar{N} = \sum_{k=0}^{K} \frac{ \dd{N_k (E > E^k_{\rm{th}})} }{ \dd{t} \dd{A} } \epsilon_k(D_k, \alpha, \tilde{E}^k_{\rm{th}}, \bar{B}),
\label{eqn:nbar}
\end{equation}
which can be then substituted into Equation \ref{eqn:likelihood} to obtain the full expression of the likelihood. 


\subsection{The associated fraction}
\label{sec:model:f}

The fraction of UHECRs associated with the source catalogue, $f$, can be calculated in this framework as a derived parameter:
\begin{equation}
f = \frac{ \sum_{k=1}^{K} F_k}{F_0 +  \sum_{k=1}^{K} F_k}.
\end{equation}
This is the expected association fraction, given the observations, and not the exact fraction of associated UHECRs in the sample. We emphasize that in addition to this summary parameter, it is also possible to calculate the explicit probability for each source-UHECR association (or equivalently, the marginal posterior distribution for the UHECR labels $\lambda_i$), as detailed in Appendix \ref{app:association}. By estimating $f$, we perform a form of embedded model comparison between the isotropic and source models within the framework of parameter estimation \citep{Kamary:2014ur}. 


\subsection{Priors}
\label{sec:model:priors}

All parameters discussed in this section and shown in Figure \ref{fig:dag} are left free when fitting the model to data. The so-called hyperparameters, or highest-level parameters, $\alpha$, $\bar{B}$, $L$ and $F_0$, have associated hyperpriors. Motivated by the desire to include our knowledge of UHECR phenomenology into the analysis and avoid giving weight to unphysical regions of the parameter space, we choose weakly informative hyperpriors for the these parameters \citep{Gelman:2006di, Simpson:2017kw, Gelman:2017hp}. Wide Gaussian priors on $\alpha$ and $\bar{B}$ reflect the lack of knowledge on these parameters and similarly for the case of $L$ and $F_0$, they reflect that we do not expect these parameters to tend to infinity. We have
\begin{equation}
\begin{split}
& P(\alpha) \sim \mathcal{N}(\mu_\alpha, \sigma_\alpha), \\ 
& P(\bar{B}) \sim \mathcal{N}(\mu_{\bar{B}}, \sigma_{\bar{B}}), \\
& P(L) \sim \mathcal{N}(0, \sigma_L / K), \\
& P(F_0) \sim \mathcal{N}(0, \sigma_F), \\
\label{eqn:priors}
\end{split}
\end{equation}
where $\mu_\alpha = 2$, $\sigma_\alpha = 3$,  $\mu_{\bar{B}} = 50$~ nG, $\sigma_{\bar{B}} = 50$~nG, $\sigma_L = 10^{43} \ \mathrm{yr}^{-1}$ and $\sigma_F = 0.01 \ \mathrm{km}^{-2}\mathrm{yr}^{-1}$. The hyperparameters have additional lower bounds: $\alpha$ at 1, $L$ and $F_0$ at 0, and $\bar{B}$ between 0 and 100~nG. The upper limit on $\bar{B}$ reflects that we do no expect such large extra-Galactic magnetic fields that would result in even nearby sources having almost isotropic arrival direction distributions. The parameters are never found on their boundaries in any of the model fits presented in this work, and a larger upper bound on $\bar{B}$ has no effect on the results.

For the analyses performed in Sections \ref{sec:sim} and \ref{sec:application}, a sensitivity analysis was performed to ensure that the main conclusions were robust to the choice of hyperprior distributions, under a range a plausible alternatives.


\subsection{Inference}
\label{sec:model:inference}

We perform inference on the model parameters for a given dataset within the framework of Bayesian probability theory. By combining the likelihood (Equations \ref{eqn:likelihood} and \ref{eqn:nbar}) and the priors (Equation \ref{eqn:priors}), we obtain an expression for the joint posterior distribution of the model parameters given the data 
\begin{equation}
P(L, \alpha, \tilde{E}, \bar{B} | \hat{E}, \hat{\omega}) \propto P(\hat{E}, \hat{\omega} | L, \alpha, \tilde{E}, \bar{B}) P(L, \alpha, \tilde{E}, \bar{B})
\end{equation}
where $P(\hat{E}, \hat{\omega} | L, \alpha, \tilde{E}, \bar{B})$ is the likelihood, given in Equation~\ref{eqn:likelihood} and $P(L, \alpha, \tilde{E}, \bar{B})$ is the joint prior distribution of the priors described in Section~\ref{sec:model:priors}.

We compute the posterior distribution numerically by generating samples from it using {\tt Stan}\footnote{\url{http://mc-stan.org}} \citep{Carpenter:2017ke}. {\tt Stan} is a software based on an implementation of adaptive Hamiltonian Monte Carlo, as outlined in \citet{Betancourt:2017vd}. We use a set of diagnostics to assess the convergence of {\tt Stan} to the target posterior distribution. These include the number of effective (or uncorrelated) samples, $n_{\mathrm{eff}}$, and the Gelman-Rubin statistic, $\hat{R}$, to monitor the mixing of separate chains \citep{Gelman:1992zz}. We require $n_{\mathrm{eff}} > 1000$ and $\hat{R} < 1.1$ with no divergent transitions for all model parameters in the analyses presented below. 

We also perform model checking to evaluate the fit of the model to data using posterior predictive checks (PPCs, \citealt{Gelman:hv, Sinharay:2003jj, Lynch:2016cm}). This involves generating data under the assumptions of the fitted model by drawing samples form the joint posterior predictive distribution. The generated data are then compared with the observed data to verify that the inferences of the model are sensible. The method and checks are detailed in Appendix~\ref{app:ppc}. 


\subsection{Arrival directions only}
\label{sec:model:arrdir}

In order to test the significance of including the UHECR energy data into the model and to facilitate comparison with previous work, we also develop a model for only the UHECR arrival directions. This follows the model described in \citetalias{Soiaporn:2012ev}, but with some key improvements. We use the above parametrization, but as there is no energy model in this case we simply infer the width of the vMF distribution representing the UHECR deflections, $\kappa$, as a parameter instead of $\tilde{E}_i$, $\alpha$ and $\bar{B}$ (see Equation~\ref{eqn:vMFdef}). Thanks to the implementation in {\tt Stan}, we are also able to infer $\kappa$ directly instead of conditioning upon it as in \citetalias{Soiaporn:2012ev}. As we do not use a Gibbs sampling approach, the choice of priors for $L$, $F_0$ and $\kappa$ are not constrained by the need to form closed-form conditional distributions for the Gibbs steps which can be directly sampled from. This makes it much simpler to implement the model and perform sensitivity analyses. Finally, the discrete latent label parameters, $\lambda_i$, are marginalised over. This allows for better exploration of the tails of the posterior distribution and more efficient sampling, as detailed in \citet{stanman}. The likelihood of this reduced model can be written (c.f. \citetalias{Soiaporn:2012ev}, Equation 19)
\begin{equation}
P(\bm{\hat{\omega}} | \bm{L}, \kappa) = \exp \left[ -\sum_{k=0}^{K} F_k \epsilon_k(\kappa) \right] \prod_{i=0}^I \sum_{k=0}^{K} F_k P(\hat{\omega}_i | \varpi_k, \kappa).
\end{equation}


\section{Simulations}
\label{sec:sim}

\begin{figure*}
\includegraphics[width=0.6\textwidth]{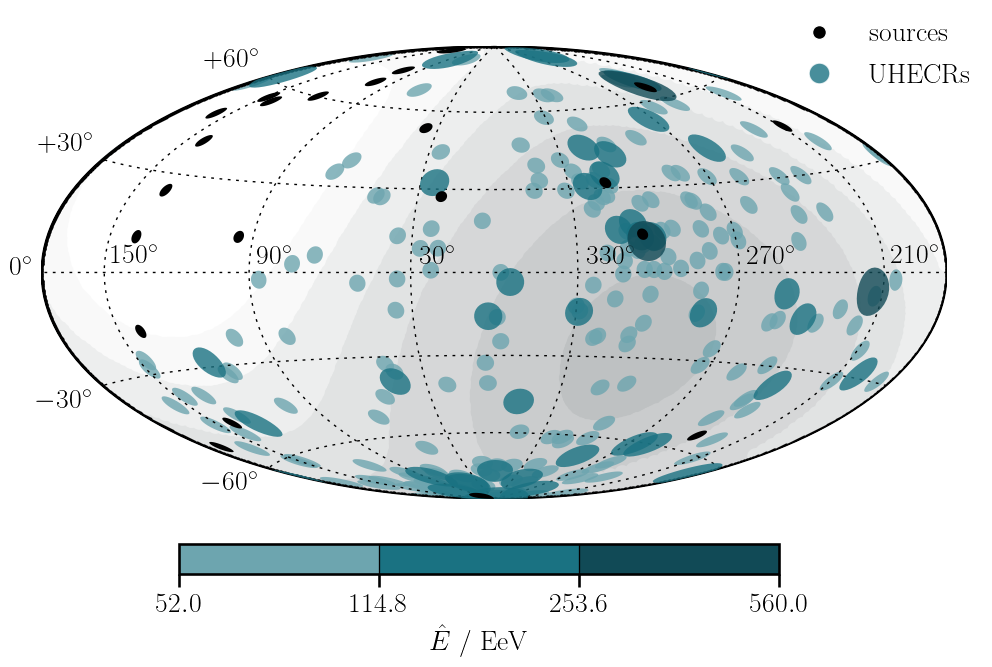}
 \caption{The simulated dataset from the generative model, shown in Galactic coordinates using the Hammer-Aitoff projection. The coloured tissots represent UHECR energies, with larger tissots corresponding to higher energy, and the greyscale shows the time-averaged exposure of the PAO. The locations of sources in the SBG catalogue are also shown by solid black tissots.}
 \label{fig:simdata}
\end{figure*}

We test our model on simulated datasets in order to verify its ability to recover the input parameters. To do this, we use the catalogue of 23 starburst galaxies (SBGs) within 250~Mpc chosen in \citetalias{Aab:2018chp} to define the source locations (see Section~\ref{sec:application:sources}). A large number of UHECRs are simulated according to the assumptions of the model with initial energies from 20~EeV, chosen as UHECRs with arrival energy below 20~EeV have negligible probability of being detected above 52~EeV, given $f_E = 0.12$. A threshold is then applied at $E_{\rm{th}} = 52$~EeV to mimic the publicly available data, yielding 249 UHECRs as shown in Figure \ref{fig:simdata}. The source spectrum is defined by $\alpha = 3$, the root mean square magnetic field strength is taken to be $\bar{B} = 20$~nG and the associated fraction of the simulated sample is set to $f = 0.5$. The observatory exposure and the detection uncertainties are modelled as described in Section \ref{sec:physics:detection}. 

\begin{figure}
\includegraphics[width=\columnwidth]{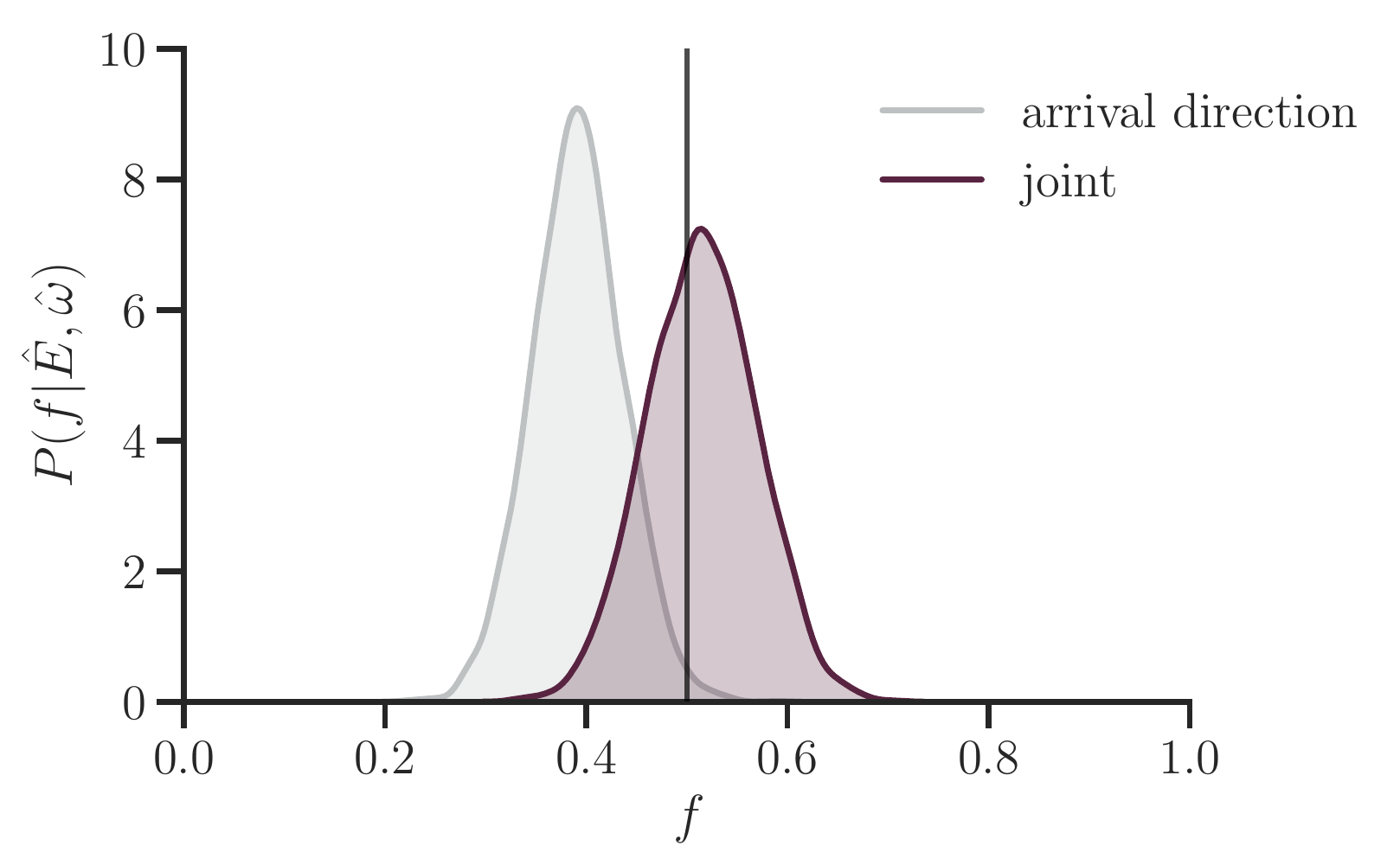}
 \caption{The marginal posterior distribution of $f$ for the case of the arrival direction only model and the full joint model. The input value of $f = 0.5$ is also shown by the black line for comparison.}
 \label{fig:energy_vs_arrival}
\end{figure}

We fit two models to the simulated sample: a model for the UHECR arrival directions only (as described in Section \ref{sec:model:arrdir}); and the full joint model for both UHECR energies and arrival directions. The resulting marginal posterior distributions for $f$ are shown in Figure \ref{fig:energy_vs_arrival}. We see that by ignoring the information in the UHECR energy data, we are not able to properly recover the input $f = 0.5$. The arrival direction model underestimates $f$ and is more constrained around this false value, with a narrower posterior distribution. In contrast, the joint model correctly recovers $f = 0.5$ and the wider posterior reflects the true uncertainty in this result due to the inclusion of the UHECR energy model. 

\begin{figure*}
\includegraphics[width=0.59\textwidth]{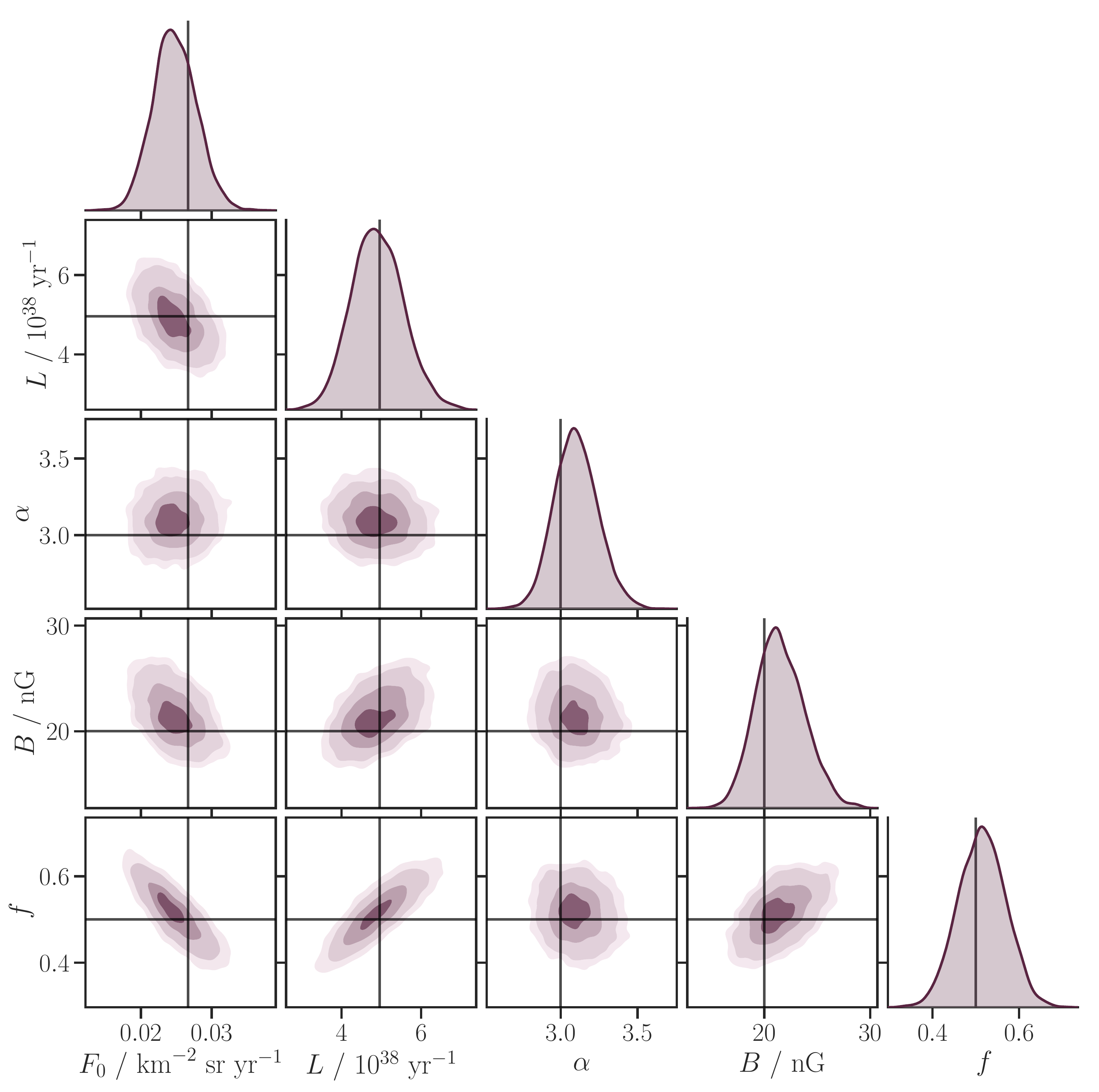}
\includegraphics[width=0.39\textwidth]{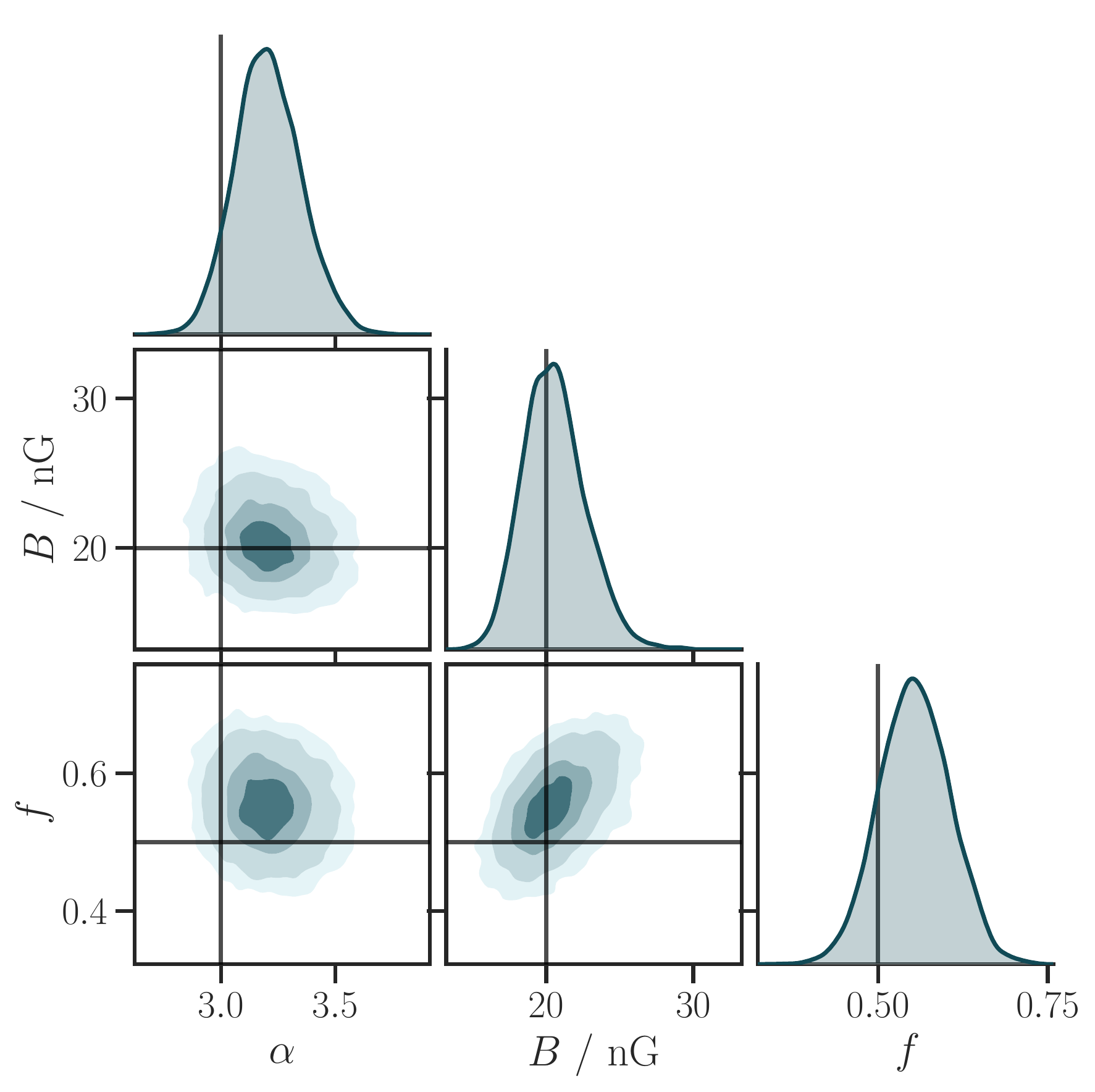}

 \caption{The joint marginal posterior distributions for the model hyperparameters from fits to the simulated datasets. The left plot shows that all input hyperparameters are inferred correctly for the case of the fit to data simulated from the generative model. The right plot shows the key physical parameters inferred from a fit to the {\tt CRPropa~3} simulated dataset. In both cases, the shaded contours show the 30, 60, 90 and 99 per cent highest posterior density credible regions and the solid lines show the input values for the simulation.}
 \label{fig:sim_corner}
\end{figure*}

It should also be noted that by not providing a physical model for the UHECR energies, we are not able to include, and thus fit for, physically interesting parameters such as $\alpha$ and $\bar{B}$. In this way, including the UHECR energies into the model is both more accurate and more informative. The joint model is able to correctly infer all input hyperparameters, as demonstrated in Figure \ref{fig:sim_corner}. This result was verified for a broad range of relevant input hyperparameters.

\subsection{CRPropa comparison}  
\label{sec:sim:crpropa}

As we recognise that our model is an approximation to the underlying physical processes involved, we also test it with the results of a Monte Carlo simulation using the publicly available {\tt CRPropa~3} code \citep{Batista:2016cb}. We use this tool to simulate the energy losses and deflections of protons more realistically, but otherwise set up an identical simulation to that described above. We include all available energy loss mechanisms due to interactions of protons with both the CMB and the extra-Galactic background light, with photomeson production simulated using the SOPHIA code \citep{Mucke:2000hr}. The magnetic field is implemented as a turbulent field with a Kolmogorov power spectrum on length scales of 1--3.135~Mpc, giving $l_{\rm{c}} = 1$~Mpc to match the assumptions described in Section \ref{sec:physics:Bfield}. We define ${\bar{B} = 20}$~nG on a 3D grid with 100~kpc spacing and 500 points in each dimension, resulting in a cube of side 50~Mpc which is repeated in space to cover the required volume. UHECRs are simulated from and tracked down to a minimum energy of 20~EeV. Again, 20~EeV is chosen as with an energy detection uncertainty of $f_E = 0.12$, UHECRs with arrival energy below 20~EeV have negligible probability of being detected above 52~EeV and thus affecting the final sample. We set up an observer sphere at the Earth with a radius of 1~Mpc (in its current implementation, {\tt CRPropa~3} requires a large observer radius in order to recover large numbers of events with reasonable computing resources). In order to offset the effect of shorter trajectories on the magnetic deflections of protons in the simulation, we also shift all sources to be 1~Mpc further away, although the impact of this on the end result is minimal. 

Despite these differences, using our model to fit the {\tt CRPropa~3} dataset we find that the marginal posterior distributions for $\alpha$ and $B$  are consistent with the input values, as shown in Figure \ref{fig:sim_corner}. This demonstrates that despite its simplicity, the model used captures the essence of the underlying physics and as such is a justifiable approximation.


\section{Application to UHECR data}
\label{sec:application}

We apply the model to the publicly available data of the 231 highest energy cosmic rays detected at the PAO, as documented in \citet{Aab:2015js}. The observatory covers an area of 3000~$\mathrm{km}^2$ and consists of 1660 surface detectors, with a duty cycle of $\sim$~100 per cent, and 24 fluorescence detectors which are limited by the need to operate in clear night conditions, resulting in a duty cycle of $\sim$~13 per cent. 

As in Section \ref{sec:sim}, we compare the case of the arrival direction only model presented in Section \ref{sec:model:arrdir} and the full joint model for UHECR energies and arrival directions as presented in Section \ref{sec:model}. 


\subsection{UHECR data}

The available data were collected between 2004 January 1 and 2014 March 1, covering a $\sim$~10 year period. The sample contains 231 UHECR above a threshold energy of $E_{\mathrm{th}} = 52$~EeV. At these energies, the trigger and selection efficiency is assumed to be 100 per cent, so the exposure of the observatory is purely a function of the detector geometry and the Earth's rotation and can be modelled as described in Section \ref{sec:physics:detection}. We also take into account the relative exposures of different observation periods, due to how the observatory has grown in size over the years, as described in \citet{Abreu:2010ce}. Additionally, the different exposure of PAO to `vertical' ($\theta_i < 60^{\circ}$) and `inclined' ($60^{\circ} \leq \theta_i < 80^{\circ}$) samples is also accounted for. These exposure effects come into the likelihood in the form of different $A_i$ and $\theta_i$, as shown in Equations \ref{eqn:margfac} and \ref{eqn:likelihood}. 


\subsection{Source catalogues}
\label{sec:application:sources}

The model presented is partly defined in terms of a catalogue of potential UHECR sources, thus the choice of sources is central to the results of the fit and the eventual conclusions of such an analysis. In this work, we follow the source catalogues used in \citetalias{Aab:2018chp}, and also by several other authors, in order to allow comparison between the different statistical approaches used. These catalogues were chosen to represent nearby AGNs and starburst galaxy (SBG) populations, motivated by the observation that the rate of energy production of UHECRs is $\sim 10^{45} \ \mathrm{erg} \ \mathrm{Mpc}^3 \ \mathrm{yr}^{-1}$, which is similar to that observed in the gamma-ray band by \emph{Fermi} for AGNs and SBGs \citep{Dermer:2010iz}. 

\begin{figure*}
\includegraphics[width=0.6\textwidth]{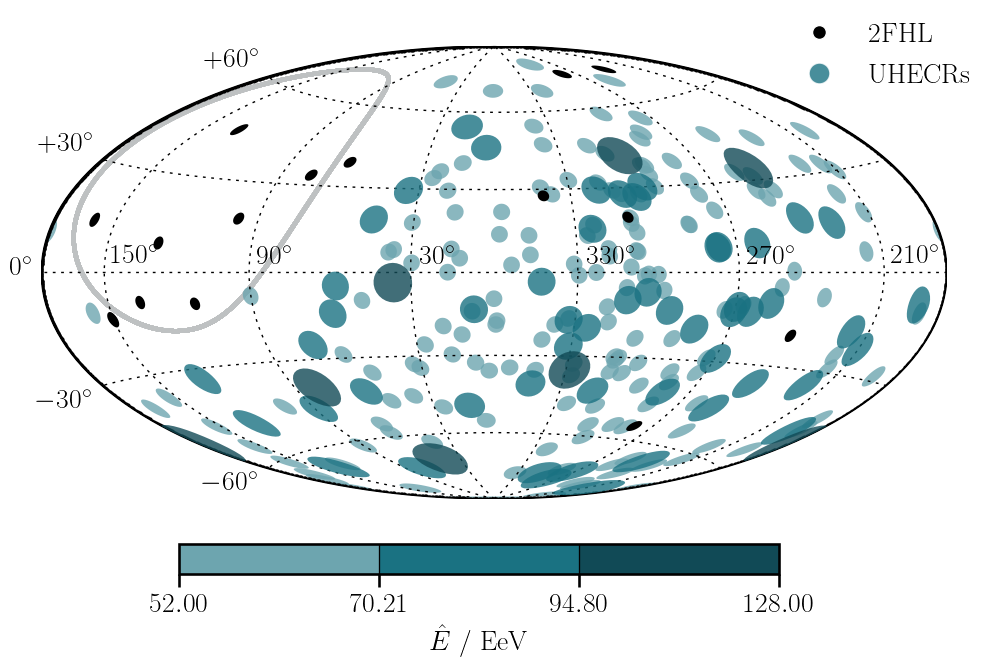}
\includegraphics[width=0.6\textwidth]{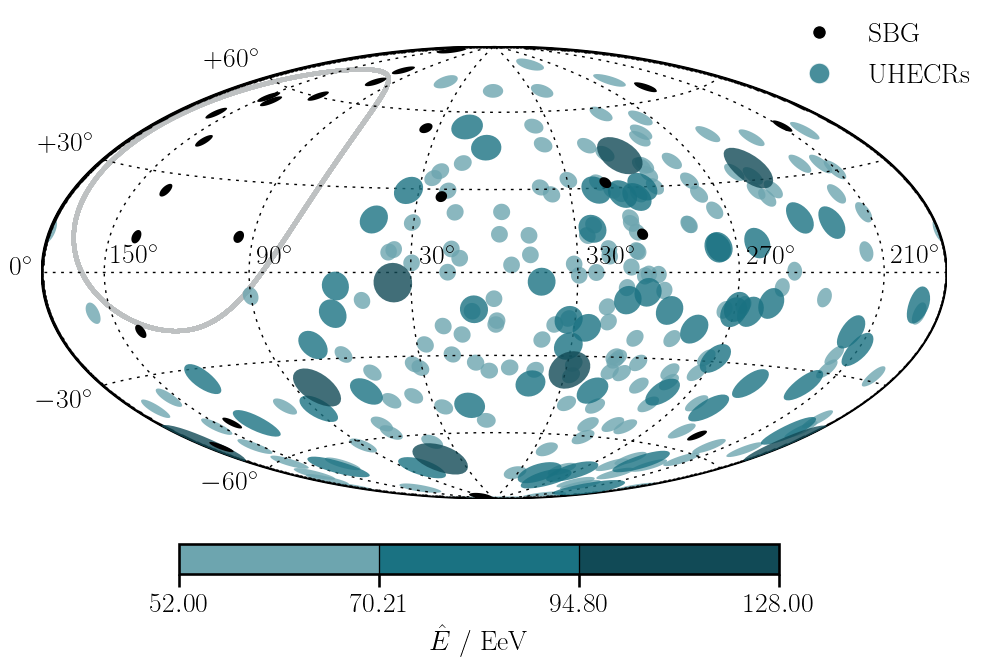}
\includegraphics[width=0.6\textwidth]{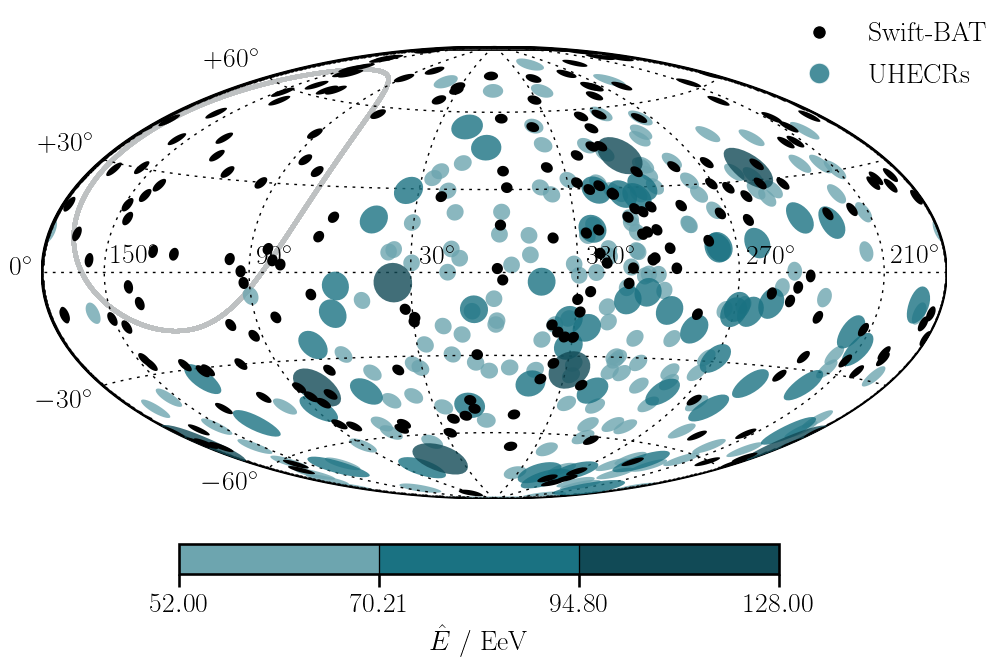}
 \caption{The source catalogues and UHECR data used in the analysis are shown in Galactic coordinates using the Hammer-Aitoff projection. The coloured tissots represent UHECR energies, with larger tissots corresponding to higher energy, and the grey solid line shows the exposure limit of the PAO. The source catalogues are also shown by solid black tissots.}
 \label{fig:sourcecat}
\end{figure*}

For completeness, we summarise the chosen catalogues here. We consider the \emph{Fermi}-LAT 2FHL catalogue of hard gamma-ray sources \citep{Ackermann:ta}, and select only radio-loud sources within a 250~Mpc radius, resulting in 17 AGN (blazars and radio galaxies). For SBGs, we take the 23 SBG which have radio fluxes $>$~0.3 Jy and are within 250~Mpc from the list of SBGs considered in a \emph{Fermi}-LAT search, reported in \citet{Ackermann:2012ih}. We also consider a flux-limited sample, the \emph{Swift}-BAT hard X-ray sources with fluxes above $13.4 \times 10^{-12}$~erg~cm$^{-2}$~s$^{-1}$ \citep{Oh:2018wzc}. These catalogues are shown along with the UHECR data in Figure \ref{fig:sourcecat}. Unlike \citetalias{Aab:2018chp}, we do not consider the 2MRS catalogue of $\sim$ 40,000 galaxies within 250~Mpc due to the large computational resources required. Under the assumptions of our model, we would expect a higher association fraction due to the sheer number of sources present. In this way a meaningful comparison with the other catalogues in our analysis would be difficult. 

Several issues have been pointed out with the specification of these potential sources. \citet{Matthews:2018laz} highlight that the nearby radio galaxy Fornax A does not appear in the 2FHL catalogue as it is an extended source, and could reasonably provide an explanation for one of the hotspots mentioned in the \citetalias{Aab:2018chp} analysis. \citet{Collaboration:2018ws} note that the exclusion of Local Group Objects (SMC, LMC, M33 and M31) is contradictory to the logic of weighting the SBG sources according to their radio fluxes, as in this case these sources would dominate the flux maps generated in \citetalias{Aab:2018chp} due to their close proximity. Considerations of statistical completeness and the selection effects of candidate source catalogues also present challenges. The specification of a physically motivated catalogue of potential UHECR sources is a non-trivial exercise and should be approached with caution in future work which attempts similar analyses. Despite the issues discussed, we proceed with the above source catalogues in the interest of a more direct comparison with previous work. 

\citetalias{Aab:2018chp} also weight the sources according to their flux in gamma-rays, radio and X-rays for the 2FHL, SBG and \emph{Swift}-BAT catalogues respectively. The details of this assumed correlation are highly model dependent, with the exact relationship between gamma-ray and UHECR flux dependent on the relevant spectra of particles and radiation in the source environment, or indeed during propagation, the UHECR composition and extra-Galactic magnetic fields \citep{Murase:2012hf, Bottcher:2013hr, Fang:2018jk}. Given that the details of this are still an open question, we instead chose the simple case of equal luminosities for the sources in our analysis.


\section{Results}
\label{sec:results}

\begin{figure*}
\includegraphics[width=0.3\textwidth]{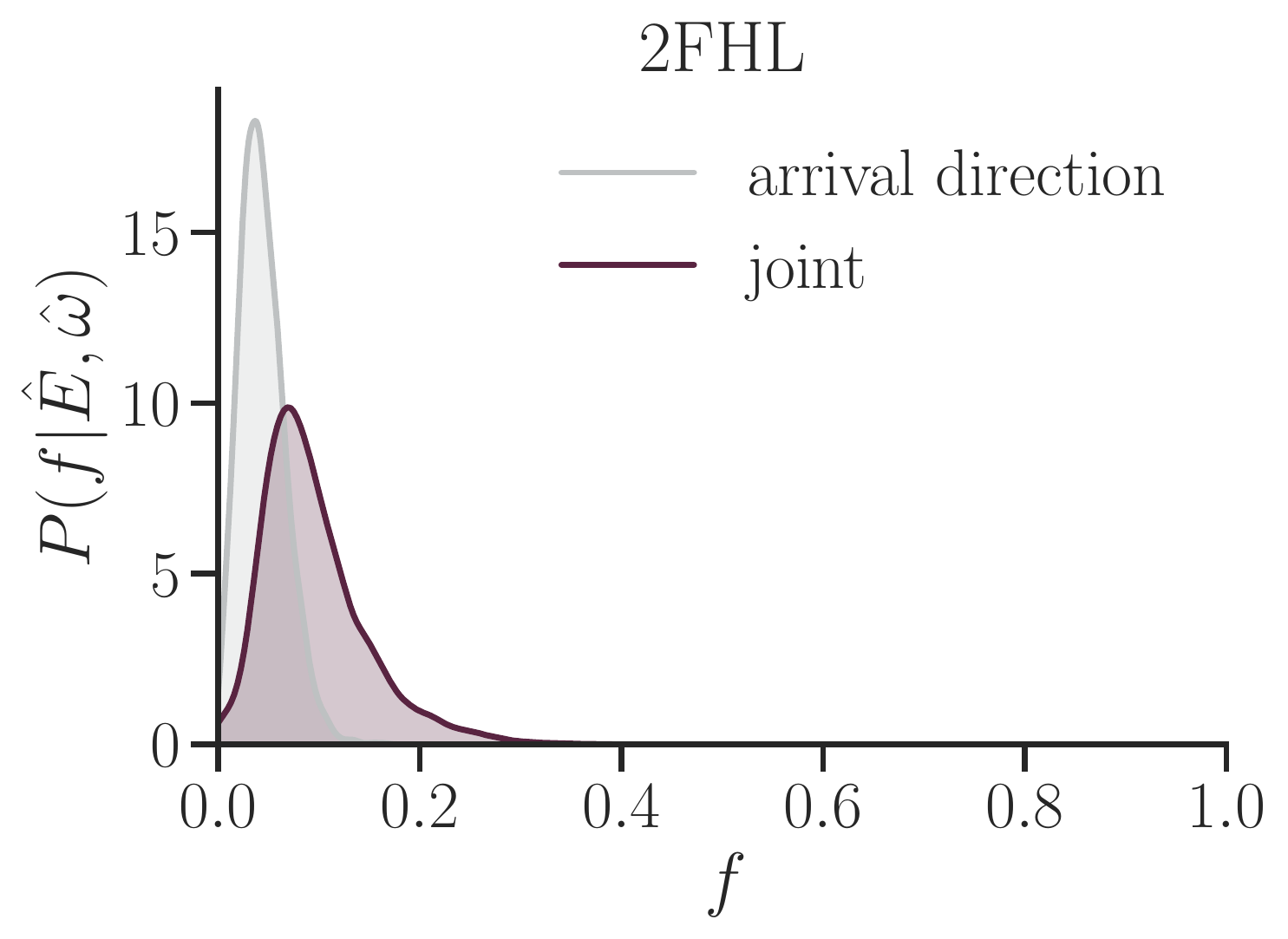}
\includegraphics[width=0.3\textwidth]{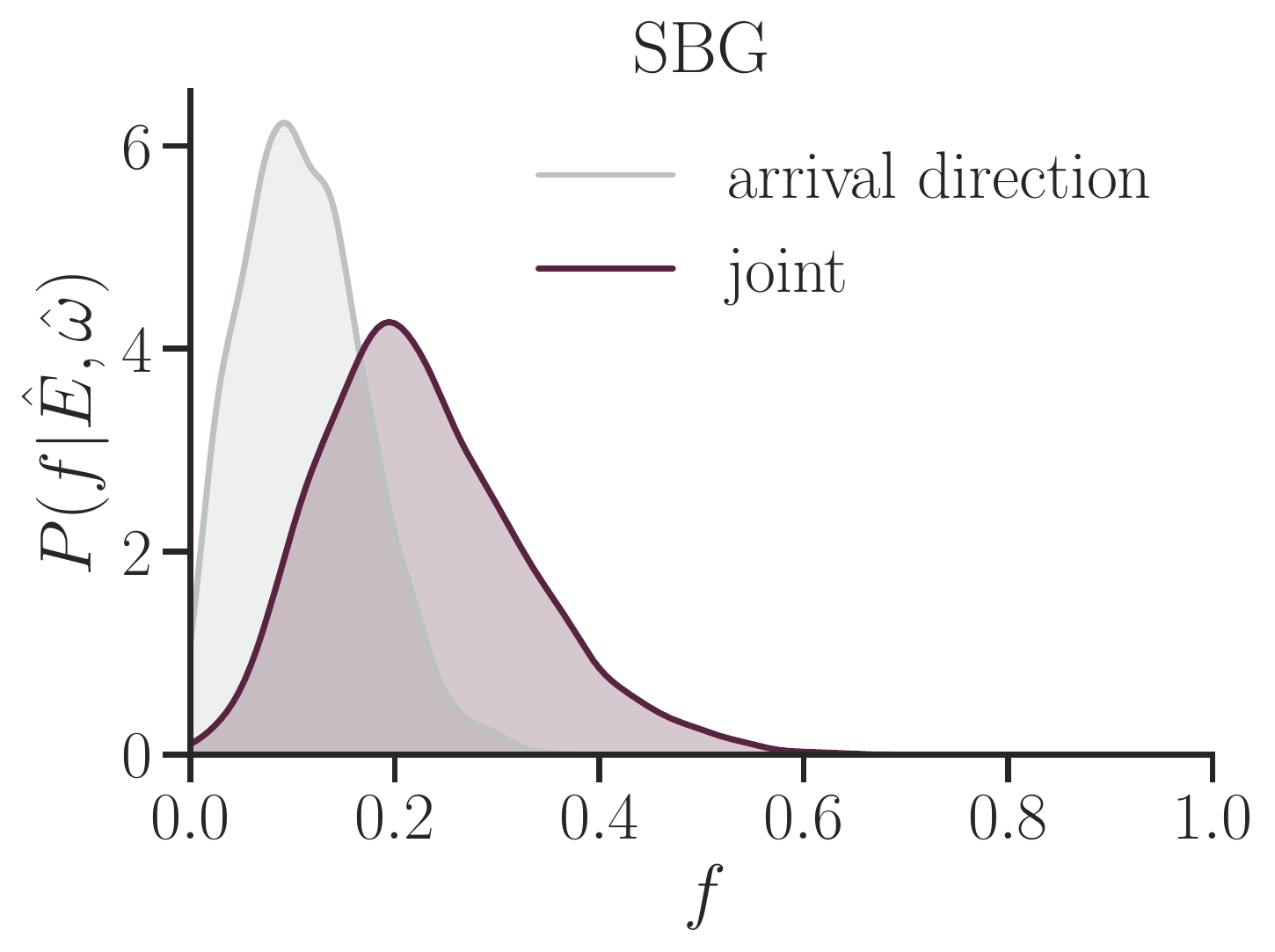}
\includegraphics[width=0.3\textwidth]{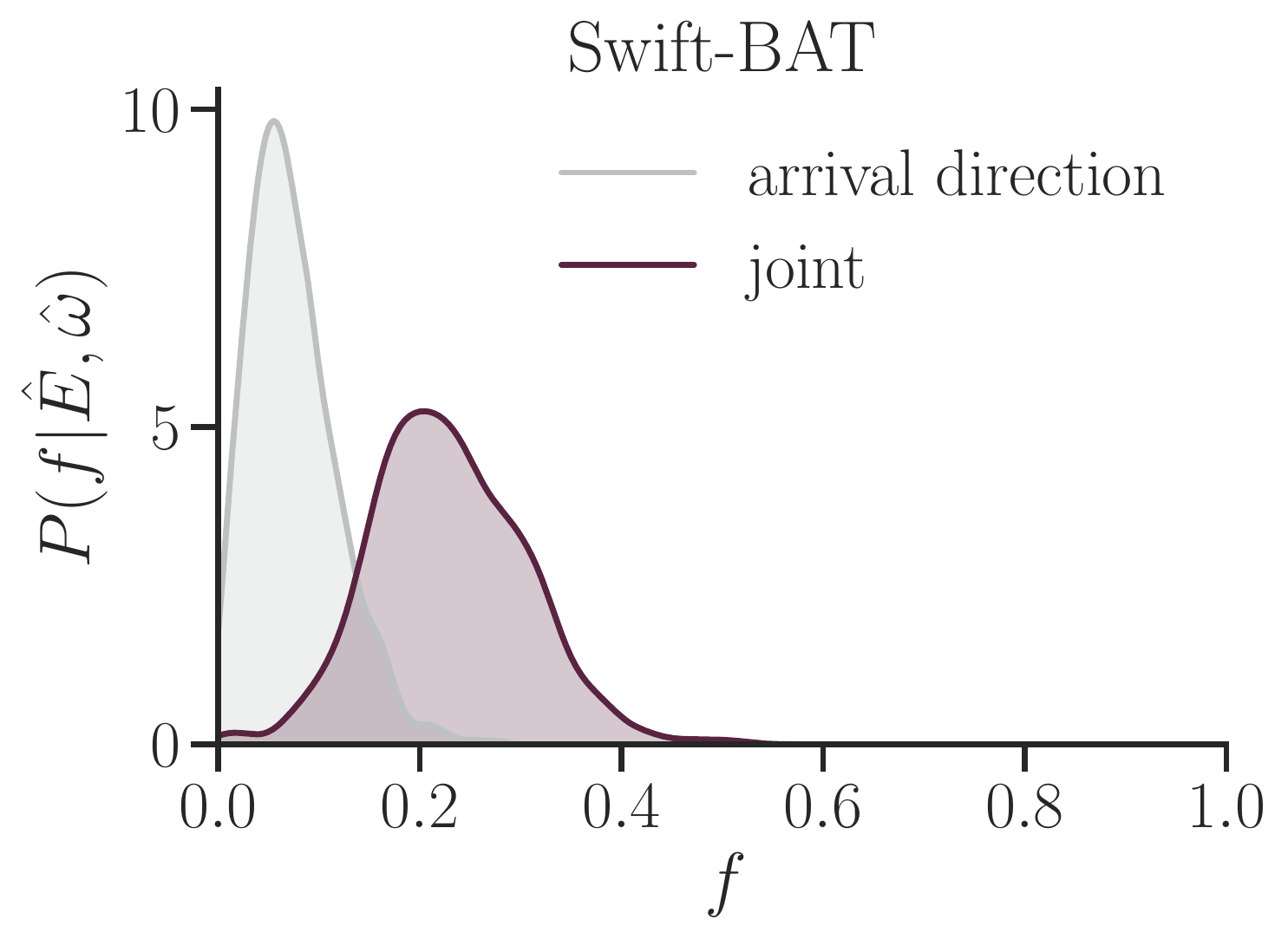}
 \caption{Marginal posterior distributions for $f$ for each of the source catalogues. In each case, we consider two different models: the arrival directions only and the joint model for UHECR energies.}
 \label{fig:f_results}
\end{figure*}

The results of the comparison between the arrival direction and joint models are shown in Figure~\ref{fig:f_results} in the form of the marginalised posterior distributions for $f$, and the corresponding 68 per cent regions of highest posterior density are given in Table~\ref{tab:f_results}. The fits of the arrival direction model find lower associated fractions compared to those of the joint model, and that $P(f | \hat{E}, \hat{\omega})$ is broader for the joint model. This is similar to what was seen for the comparison of the two models using the simulated dataset, as discussed in Section~\ref{sec:sim} and shown in Figure~\ref{fig:energy_vs_arrival}. It seems by including the energy information in the context of this model, we are able to uncover stronger associations between the candidate source catalogues and the UHECRs. We interpret the broader marginal posterior distributions as a more realistic representation of the uncertainty in the results, due to the inclusion of a model for the UHECR energy losses and reconstruction. The largest associated fractions of $\sim$~23 per cent are found for the case of the SBG and \emph{Swift}-BAT catalogues. The fact that we find roughly the same association fraction for both the SBG and \emph{Swift}-BAT catalogues demonstrates that very different catalogues can produce similar results under the assumptions of our model. The SBG catalogue is made up of 23 predominantly nearby sources, whilst the \emph{Swift}-BAT catalogue contains 213 sources, most of which are more distant. By allowing all model parameters to be free in the fit, we see that different values of the magnetic field strength allow for the density of sources in each case to explain a reasonable fraction of the UHECR observations. Both catalogues trace the local matter distribution in different ways. 

Our results are broadly consistent with the most recent results of the PAO for single catalogues against isotropy (Section~4.2 in \citetalias{Aab:2018chp}), which finds associated fractions of $7 \pm 4$, $10 \pm 4$ and $7^{+4}_{-3}$ per cent for the 2FHL, SBG and \emph{Swift}-BAT catalogues respectively, with additional systematic errors of $\sim$~0.3 per cent. We see that the results are generally closer to those of the arrival direction model, with our joint model favouring higher associated fractions in each case. Additionally, we tend to have larger uncertainties on our results which are naturally represented by the width of the marginal posterior distribution. It is important to note that larger uncertainties do not necessarily mean that our method is less constraining, but rather reflects that it consistently takes into account the uncertainties at each level of the model hierarchy. By reducing the number of assumptions we make with regards to the model in order to make it more physical, we naturally introduce uncertainties into the problem. For example, by fixing $\alpha$ or $\bar{B}$, we obtain narrower marginal posterior distributions for $f$, but the apparent constraining power is simply a result of making stronger assumptions. As we are able to infer physical parameters, there is no need to fix or scan over any of the parameters in the model.

A direct comparison between our results and the work of \citetalias{Aab:2018chp} is not possible for a number of reasons. Firstly, the PAO collaboration have access to the full, more recent UHECR dataset with a total exposure of $\alpha_{\rm{T}} = 89,720$~km$^2$~sr~yr, and also to lower energy events, with their analysis considering 5514 events down to 20 EeV. Secondly, we consider a pure proton model, whereas they compute the attenuation weights for a mixed composition based on the results of a joint analysis of the UHECR spectrum and composition \citep{Aab:2016zth}, which are then included in the generation of the presented flux maps. Finally, they also assume a range of fixed source spectral indices ($\alpha = \{-1.5, 1, 2\}$) in the context of different scenarios which are chosen to represent a range of compositions and maximum rigidities at the sources. In this way, there are many factors that are different between the two analyses in addition to the statistical method employed and so the results should be interpreted accordingly.

\begin{figure*}
\includegraphics[width=0.6\textwidth]{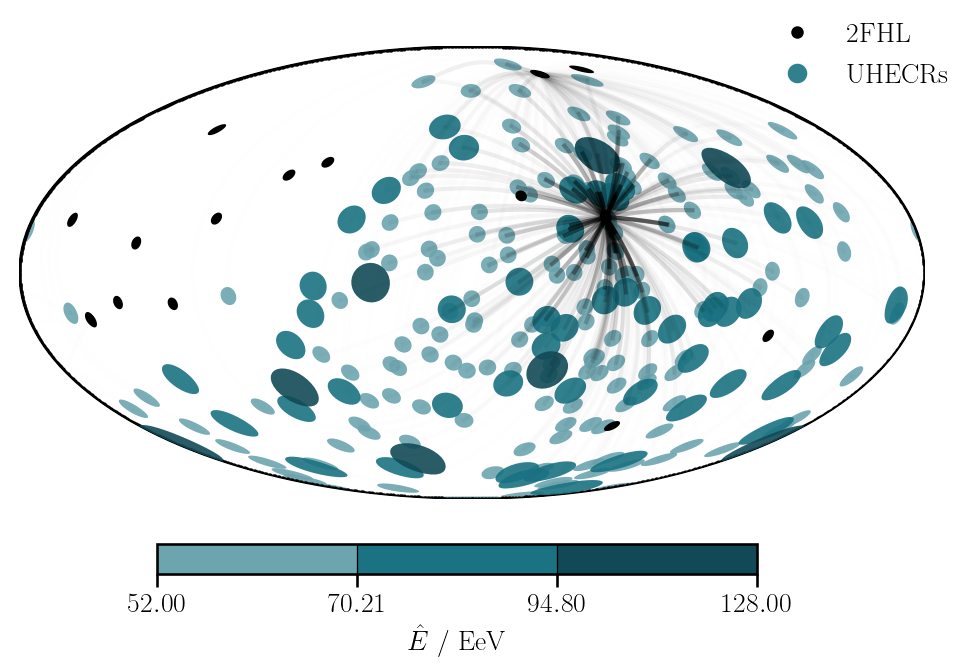}
\includegraphics[width=0.6\textwidth]{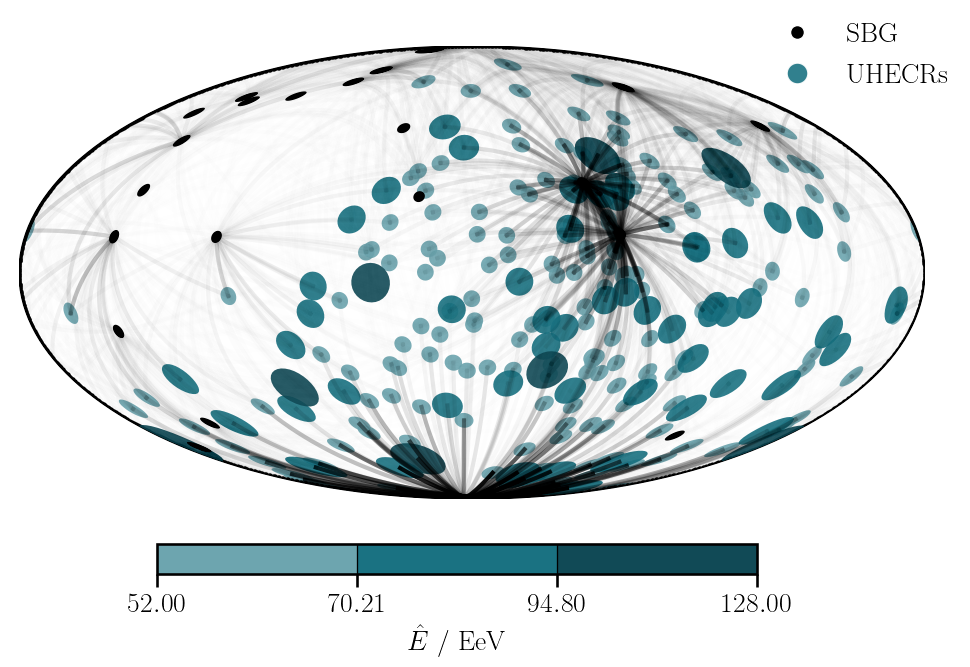}
\includegraphics[width=0.6\textwidth]{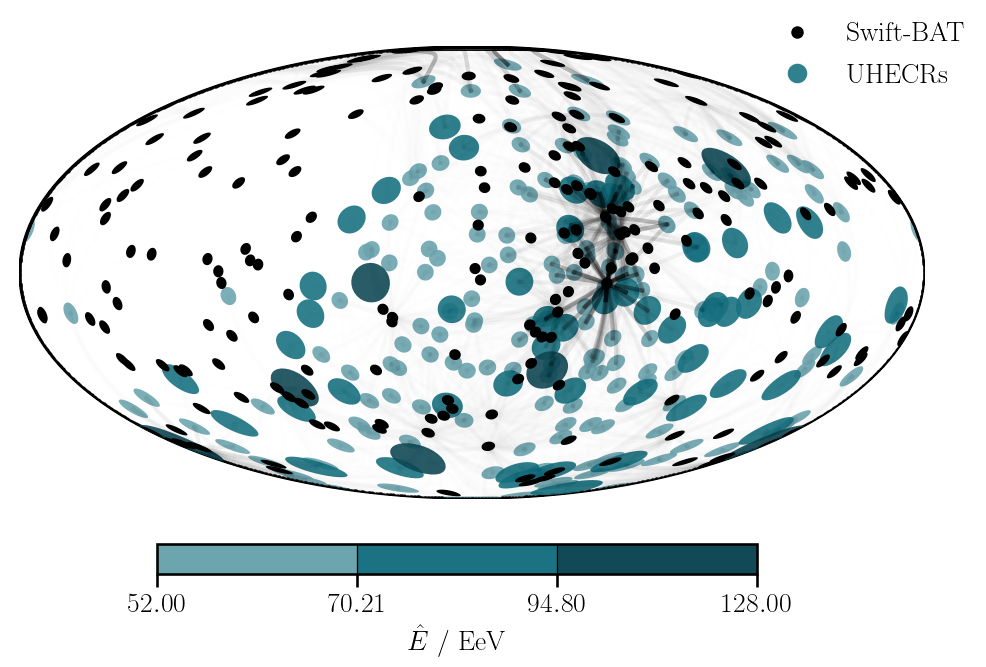}
 \caption{The association probabilities of source-UHECR pairs resulting from the fit of the joint model are plotted together with the UHECR data for the different source catalogues. The lines are shaded according to the association probabilities (normalised to the maximum) and are plotted for the case of ${P(\lambda_i = k | \hat{E}, \hat{\omega}) > 0.001}$. The dominant sources are described in the text. As in Figure~\ref{fig:sourcecat}, the sky map is shown in Galactic coordinates using the Hammer-Aitoff projection, the coloured tissots represent UHECR energies, with larger tissots corresponding to higher energy, and the source catalogues are also shown by solid black tissots.}
\label{fig:association}
\end{figure*}

In order to illustrate the fits and further analyse the results for $f$, we compute the association probabilities of each source-UHECR pair as described in Appendix~\ref{app:association} and summarise the results in Figure~\ref{fig:association}. As expected for equal source luminosities, the associations are dominated by the nearest sources. For the 2FHL catalogue, Centaurus~A dominates with a small contribution from M~87. For the SBGs, NGC~4945, M~83 and NGC~253 have the strongest associations with small contributions from almost all of the catalogue reflecting the higher value of $f$. In the case of the \emph{Swift}-BAT catalogue, Centaurus~A and the Circinus galaxy dominate. Here the large number of sources mean that the contribution to the higher $f$ value mostly comes from many low-probability associations. Figure~\ref{fig:association} also allows us to visualise the implication of the value of $\bar{B}$ resulting from the fit. We can see that for the 2FHL catalogue, the large tail in $P(\bar{B} | \hat{E}, \hat{\omega})$ allows for low probability associations with low energy UHECRs that are far from Centaurus~A and M~83. The larger $\bar{B}$ found for the SBG case is represented by the stronger association probabilities for more deflected UHECRs whereas the smaller $\bar{B}$ found for the \emph{Swift}-BAT case is shown by the preference for smaller deflections.

\begin{figure}
\includegraphics[width=\columnwidth]{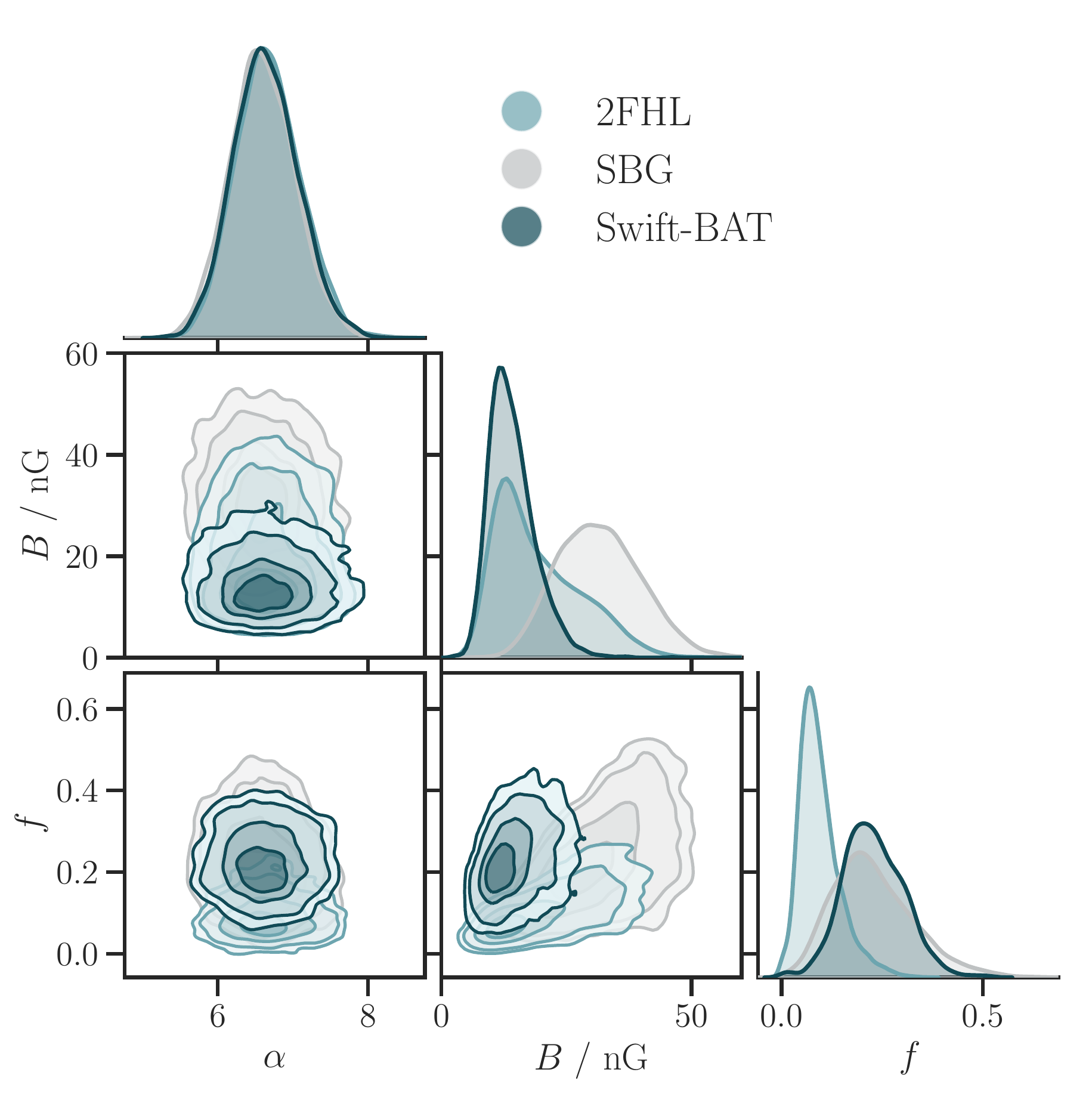}
 \caption{Joint marginal posterior distributions of $\alpha$, $\bar{B}$ and $f$ for each of the candidate source catalogues. The shaded contours enclose the 30, 60, 90 and 99 per cent highest posterior density credible regions.}
 \label{fig:param_results}
\end{figure}

For the joint model fits, the marginal posterior distributions of key parameters are summarised in Figure~\ref{fig:param_results}. For each catalogue, we see that the fit favours a very steep energy spectrum of $\alpha \sim 6.5$. This can be understood by the fact that with the majority of UHECRs being associated with the background component of the model, the $\alpha$ resulting from the fit really represents the arrival energy spectrum and not that of the sources, and is therefore the same for each catalogue. The large $\alpha$ is needed to fit the arrival spectrum due to the lack of an exponential cutoff in our model for both the injection spectrum and the arrival spectrum. All fits generally favour $\bar{B} \sim 10 - 30$~nG, with a longer tail out to $\sim50$~nG for the 2FHL catalogue. These values correspond to deflections on the scale of $1.15^{\circ} (\bar{B} / 1 \ \mathrm{nG})$ for a UHECR of 100~EeV travelling a distance of 10 Mpc (see Equation~\ref{eqn:deflection}). The deflections are not small for lower energy UHECRs and the approximation used is not strictly valid, but reflects the underlying relationship of the relevant parameters with the data. 

Given the simplicity of our model, we do not attempt to draw direct astrophysical conclusions from the fit values of $\alpha$ and $\bar{B}$, but it is important to demonstrate that the described framework is able to include such parameters and these results can inform on future extensions to the propagation model which would allow for a more physical interpretation. An obvious advantage of the framework is that it is possible to simultaneously perform inference on a magnetic field model and a UHECR model avoiding the need to fully specify one in order to interpret the other. This is one of the main goals of the IMAGINE consortium\footnote{\url{https://www.astro.ru.nl/imagine}}, which aims to bring together data from the fields of Galactic science, large scale structure formation cosmic rays and to build a consistent picture of the phenomenology using Bayesian methods \citep{Boulanger:2018ug}.

The model for the background component assumes no UHECR energy losses and thus has the flexibility required to provide a good fit to the data. A more physical model in which the background UHECRs are modelled as coming from beyond 250~Mpc was also considered and tested against the data. This model is much more constraining, requiring UHECRs which cannot be explained by the catalogued sources to be of lower energies, due to their more distant origin, and was unable to provide a good fit to the data for any of the source catalogues considered. This can be understood by considering the location of several of the highest energy UHECRs, shown in Figure~\ref{fig:sourcecat}, which are either far from potential sources (2FHL, SBG), or the sources which seem `nearby' on the celestial sphere are in fact distant (>~50~Mpc for \emph{Swift}-BAT). We see in Figure~\ref{fig:association} that we do not have any source-UHECR associations for three of the highest energy UHECRs. Within the context of our model, UHECRs of these energies (> 100 EeV) are not strongly deflected in magnetic fields and cannot be from distant sources (see Figure~\ref{fig:P_x_z}). In this way, the catalogues proposed cannot explain the UHECR data observed and thus the fit favours the more flexible isotropic background component, even if large deflections are considered. 

We use PPCs to assess the ability of the joint model to fit the data and the results are shown in Appendix~\ref{app:ppc}. We find reasonable fits for all cases, with possible indications of extensions to the spectral and luminosity function modelling being able to better represent the data.

\bgroup
\def\arraystretch{1.5} 
\begin{table}
\centering
\caption{The results for $f$ for both the arrival direction and joint models. The results are given as the mean value with the lower and upper bounds representing the 68 per cent region of highest posterior density. }
\label{tab:f_results}
\begin{tabular}{lll}
& \multicolumn{2}{c}{$f$ \%} \\
\cline{2-3}
& Arrival direction & Joint \\
\hline
2FHL & $4.4^{+1.7}_{-2.6}$  & $9.5^{+2.4}_{-5.9}$ \\ 
SBG & $11.2^{+4.6}_{-7.6}$ & $22.7^{+6.6}_{-12.4}$ \\
\emph{Swift}-BAT & $7.4^{+2.6}_{-5.4}$ & $22.8^{+6.6}_{-8.0}$ \\
\hline
\end{tabular}
\end{table}
\egroup


\section{Conclusions}
\label{sec:conclusion}

We present a hierarchical model for the UHECR energies and arrival directions, enabling a joint fit of a phenomenological model to the UHECR observations. 

Through calculations and simulations within the framework of the model, we demonstrate the impact of including UHECR energy data and detection effects on the physical interpretation. Modelling reconstruction uncertainties of 12 percent for the UHECR energies results in an effective extension of the GZK horizon due to the misidentification of lower energy events in a UHECR sample. The size of this effect depends on the shape of the source spectrum, the model for the UHECR energy losses and the distribution of sources, but is non-negligible in many cases with $\sim30$ per cent of a selected UHECR sample coming from beyond 250~Mpc for the example presented in Figure~\ref{fig:Edet_effect}. We also include the arrival direction reconstruction uncertainty of $1^{\circ}$ and show the influence of the detection effects on the particle-specific effective GZK horizon. Again, the distribution of possible source locations of the UHECRs is considerably extended, with the energy uncertainty being mostly responsible. 

By comparing fits of simulated data with the joint model and a simpler model for only the UHECR arrival directions, we show that including the UHECR energy information is essential in order to recover the input parameters. We interpret these results as a strong motivation for explicitly modelling the individual UHECR energies in future comparisons of candidate source populations and UHECR data. The joint model also enables to fit for the source spectral index and the RMS magnetic field strength and can reasonably recover the input parameters of simulations run using {\tt CRPropa~3}, showing that the model is a justifiable approximation capable of inferring physical parameters. 

The model is applied to the publicly available data of 231 UHECR with $\hat{E} > 52$~EeV detected by the PAO and the 2FHL, SBG and \emph{Swift}-BAT catalogues of potential sources. We find the largest association fractions of $22.7^{+6.6}_{-12.4}$ and $22.8^{+6.6}_{-8.0}$~per~cent for the SBG and \emph{Swift}-BAT catalogues respectively. The fact that very different source distributions can give similar results motivates extension of the model in order to test more constraining scenarios against the data. Whilst our model is able to fit the data well, several of the highest energy cosmic rays in the sample are left without strong source associations. This implies that the chosen catalogues cannot fully explain the observations under the assumptions of the model, and can inform on the choice of future candidate source catalogues. 

Both the physical model and the formalism presented here can be extended to include more complex acceleration and propagation models. Additionally, we plan to include the available TA data \citep{Abbasi:2014hi} and detector exposure into this framework in future work and to account for the ${\sim14}$ per cent systematic uncertainties on the UHECR energy scale.

Results from the PAO favour a heavier composition at the highest energies over pure protons \citep{Aab:2016ci}. An obvious extension of our model would be to introduce the possibility of a mixed composition. This could be achieved by modelling the composition as an additional latent mark on the production, resulting in a marked Poisson process with three marks: $\{\lambda_i, \tilde{E}_i, C_i \}$. The simplest case would be to approximate the mark distribution for the $C_i$ as a mixture model over five representative elements at the source, e.g. over H, He, N, Si and Fe, as is typically done when modelling the composition. The weights of the mixture model could be left as free parameters to directly infer the source properties based on the data. A propagation model then needs to be specified for the composition-dependent energy losses and deflections. For the deflections, a straightforward option would be to consider Equation~\ref{eqn:deflection} multiplied by the initial charge of the UHECRs, $Z$, and the energy losses could also be approximated to be purely a function of the initial composition of the UHECRs. The reconstruction of the composition from air shower data can be treated using Gumbel distributions \citep{Poincare:wt} which can be used to model the distribution of extreme values ($X_{\rm{max}}$ in this case) together with the parameterisation described in \citet{DeDomenico:2013fl}. 

However, a realistic physical model needs to account for the changing composition during the UHECR propagation due to photodisintegration and the effect of this on the energy losses and magnetic deflections. For this reason we model the UHECRs as protons here in order to demonstrate the use of the hierarchical model and only highlight the capability of the model to include the UHECR composition as a possibility for future work. An interesting avenue to explore in this case would be to interface a Monte Carlo simulation code for the UHECR propagation, such as {\tt CRPropa~3}, with the hierarchical model and statistical framework described here.


\section*{Acknowledgements}
We thank J. M. Burgess and C. Fuglesang for useful discussions. This work made use of NumPy~\citep{numpy}, SciPy~\citep{scipy}, Astropy~\citep{astropy:2018}, Matplotlib~\citep{Hunter:2007ouj}, Seaborn~\citep{seaborn}, h5py \citep{collette_python_hdf5_2014} and the High Performance Computing facilities at the PDC center of the Swedish National Infrastructure for Computing. FC was supported by the Alexandra \& Bertil Gylling Foundation.  




\bibliographystyle{mnras}
\bibliography{hierarchical_uhecr}



\appendix


\section{Derivation of the likelihood}
\label{app:likelihood}

Following the ideas introduced in Section~\ref{sec:model:likelihood}, we proceed by starting at the top of the model hierarchy with the source population. The total population can be modelled as a superposition of homogeneous Poisson point processes from each component. UHECRs are produced in a marked Poisson point process, which means that they are produced with a set of associated properties, referred to as `marks': $\{\lambda_i, \tilde{E}_i\}$. The $\lambda_i$ are integer valued labels associating UHECRs $i$ to a source components $k$. Their categorical mark distribution is given by 
\begin{equation}
P(\lambda = k | F_k) = \frac{F_k}{\sum_{j = 0}^{K} F_j}.
\end{equation} 
These labels implicitly define the initial directions of the UHECRs, since they are produced at the source position. The $\tilde{E}_i$ are distributed according to the source power law spectrum, characterised by $\alpha$ and normalised as described in Section \ref{sec:physics:acc}. The resulting mark probability distribution for the latent energy marks, $\tilde{E}_i$, is thus a Pareto distribution \citep{Arnold:2015uz}
\begin{equation}
P(\tilde{E}_i | \alpha) = \frac{\alpha - 1}{E_{\rm{min}}} \left( \frac{\tilde{E}_i}{E_{\rm{min}}} \right)^{-\alpha}.
\end{equation}

UHECR propagation involves both energy losses and magnetic deflections, as described in Section~\ref{sec:physics:propa}. The arrival energies $E_i$ are calculated directly from the $\tilde{E}_i$ by solving Equation~\ref{eqn:E_loss} for a given source distance, $D_k$, and UHECR redshift, $z$. In this way, $P(E_i | \tilde{E}_i, D_k)$ is a delta function
\begin{equation}
P(E_i | \tilde{E}_i, D_k) = \delta \left[ E_i - E(\tilde{E}_i, D_k)  \right].
\end{equation}
For the background component with $k = 0$, there are no energy losses and $E_i = \tilde{E}_i$, which is the equivalent of the above statement with $D_0 = 0$~Mpc. Magnetic deflections are modelled using the vMF distribution (defined in Section~\ref{sec:physics:detection}) centred on the source locations
\begin{equation}
P(\omega_i | \varpi_k, D_k, \tilde{E}_i, \bar{B}) = \frac{ \kappa_{k,i} }{ 4 \pi \sinh( \kappa_{k,i}) } e^{\kappa_{k,i} \omega_i . \varpi_k},
\label{eqn:vMFdef}
\end{equation}
where $\kappa_{k,i} = \kappa_{k,i}(D_k, \tilde{E}_i, \bar{B})$, as given by $\theta_\mathrm{rms}$ in Equation~\ref{eqn:deflection}, which is then converted to the dimensionless vMF parameter using Equation~\ref{eqn:convMF}. 

We now proceed to UHECR detection and reconstruction. The form of this component of the likelihood is given in Equation~\ref{eqn:inhomopp}. We first focus on the rate of the process. It is possible to marginalise over the latent UHECR arrival directions and find an analytic solution to the integral such that \citepalias{Soiaporn:2012ev}
\begin{multline}
P(\hat{\omega}_i | \varpi_k, D_k, \tilde{E}_i, \bar{B}) = \int \dd{\omega_i} P(\hat{\omega} | \omega_i) P(\omega_i | \varpi_k, \kappa_{k,i}) A_{\perp}(\omega_i, t_i) \\ 
\approx A_i \cos(\theta_i) \int \dd{\omega_i} \ P(\hat{\omega} | \omega_i) P(\omega_i | \varpi_k, \kappa_{k,i}) \\
= A_i \cos(\theta_i)\begin{cases} \frac{\kappa_{k,i} \kappa_{\rm{d}}}{ 4 \pi \sinh( \kappa_{k,i}) \sinh( \kappa_{\rm{d}}) } \frac{\sinh(| \kappa_d \omega_i + \kappa_{k,i} \varpi_k |)}{| \kappa_{\rm{d}} \omega_i + \kappa_{k,i} \varpi_k |} & \mbox{if } k \geq 1 \\ \frac{1}{4 \pi} & \mbox{if } k = 0 \end{cases},
\label{eqn:margfac}
\end{multline}
where $A_{\perp}$ is the effective area of the detector at time $t$, $\theta_i$ is the detected UHECR zenith angle and we have assumed that the angular reconstruction uncertainty, $\sigma_\omega$, is small compared to the scale over which $A_\perp$ varies in the approximation. The dependence of $\kappa_{k,i} = \kappa_{k,i}(D_k, \tilde{E}_i, \bar{B})$ has been omitted for clarity. $A_i$ denotes the area of the detector at UHECR arrival time $t_i$. This changes slowly as the size of a detector grows during its construction, as for the different periods described in \citet{Abreu:2010ce}. The resulting expression for the Poisson rate is given in Equation \ref{eqn:likelihood}.

The expected number of events, $\bar{N}$, is given by the sum over the contributions from all possible sources, as described in Section \ref{sec:model:likelihood}. The effective exposure, $\epsilon_k $, is indirectly dependent on the source UHECR energy distribution, the source distance and the magnetic field strength through the dependence of the vMF distribution on $\kappa_{k,i} = \kappa_{k,i}(D_k, \tilde{E}_i, \bar{B})$. As $\kappa_{k,i}$ is different for each UHECR, and we do not know the source-UHECR associations, $\tilde{E}_i$ or $\bar{B}$ a priori, we approximate the spread of the vMF by considering an expected $\kappa$ for each source, $\bar{\kappa}_k$. We have
\begin{equation}
\epsilon_k(\bar{\kappa}_k) =  \begin{cases}  \int \rm{d} \omega \ P(\omega | \varpi_k, \bar{\kappa}_k) \epsilon(\omega) & \mbox{if } k \geq 1 \\ 
\alpha_T / (4 \pi) & \mbox{if } k = 0 \end{cases},
\end{equation}
which is weakly dependent on $\bar{\kappa}_k$, particularly in the limit of $\bar{\kappa}_k \gg 1$. 

In order to summarise the distribution of energies at the source, we consider the median of the source power law distribution: $\tilde{E}_m = 2^{1 / (\alpha - 1)} \tilde{E}_{\rm{th}}$. Consequently, $\bar{\kappa}_k = \bar{\kappa}_k(D_k, \alpha, \tilde{E}_{\rm{th}}, \bar{B})$ and is independent of the UHECR index, $i$. Bringing this result together with Equation \ref{eqn:flux_gzk} and summing over all source components, we arrive at Equation \ref{eqn:nbar}.


\section{Source-UHECR association}
\label{app:association}

As described in Section~\ref{sec:model:likelihood}, UHECRs are modelled as having integer labels, $\lambda_i$, which denote their sources. Having marginalised out the $\lambda_i$ in the likelihood (Equation \ref{eqn:likelihood}), meaning that the $\lambda_i$ are not explicitly sampled during inference, we now show how to recover the marginal posterior distribution for $\lambda_i$, as described for a similar problem in \citet{stanman}. The likelihood function can be expressed as
\begin{equation}
\begin{split}
P(\hat{E}, \hat{\omega} | L, \alpha, \tilde{E}, \bar{B})& \propto \prod_{i=0}^N \sum_{k=0}^{K} P(\lambda_i = k, \hat{E}_i, \hat{\omega_i} | L, \alpha, \tilde{E}, \bar{B}) \\
& \propto \prod_{i=0}^N \sum_{k=0}^{K} P_{k, i},
\end{split}
\end{equation}
where the inner term, $P_{k, i}$, is the joint probability for $\lambda_i = k$ and the observed data, given the model parameters. During sampling, assuming convergence, the values for the model parameters are drawn from the joint posterior distribution, $P(L, \alpha, \tilde{E}, \bar{B} | \hat{E}, \hat{\omega})$. At each iteration, the associated $P_{k,i}$ is calculated. Averaging this term over the iterations gives an estimate of the unnormalised marginal posterior for $\lambda_i = k$
\begin{equation}
P(\lambda_i = k | \hat{E}, \hat{\omega}) \propto q(\lambda_i = k | \hat{E}, \hat{\omega}) = \frac{1}{S} \sum_{s=0}^S P^s_{k, i},
\end{equation}
where $S$ is the number of iterations, or posterior samples. We can then normalise this by dividing by the sum over all possible associations of the $i^{\rm{th}}$ UHECR
\begin{equation}
P(\lambda_i = k | \hat{E}, \hat{\omega}) = \frac{q(\lambda_i  = k | \hat{E}, \hat{\omega})}{\sum_{l = 0}^{K} q(\lambda_i = l | \hat{E}, \hat{\omega})}, 
\end{equation}
thus recovering the normalised marginal posterior distribution for the individual UHECR labels $\lambda_i$.


\section{Posterior predictive checks}
\label{app:ppc}

In order to assess the ability of our model to fit the data, we perform PPCs as introduced in Section~\ref{sec:model:inference}. The posterior predictive distribution for this dataset is defined as
\begin{equation}
P(\hat{E}_{\rm{rep}}, \hat{\omega}_{\rm{rep}} | \hat{E}, \hat{\omega}) = \int \dd{\Theta} P(\hat{E}_{\rm{rep}}, \hat{\omega}_{\rm{rep}} | \Theta) P(\Theta | \hat{E}, \hat{\omega}), 
\end{equation}
where $\Theta$ is the set of all model parameters and $\hat{E}_{\rm{rep}}$ and $ \hat{\omega}_{\rm{rep}}$ are the replicated data that we wish to generate. We draw $\hat{E}_{\rm{rep}}$ and $\hat{\omega}_{\rm{rep}}$ from this distribution and compare with the observed $\hat{E}, \hat{\omega}$ in order to evaluate the how well the model represents the data. Systematic differences between the replicated and observed data are an indicator of model misfit. In this way, in addition to assessing the quality of a fit, PPCs can be used as a tool to indicate in which ways to extend a model in order to better represent the data. 

\begin{figure*}
\includegraphics[width=0.95\columnwidth]{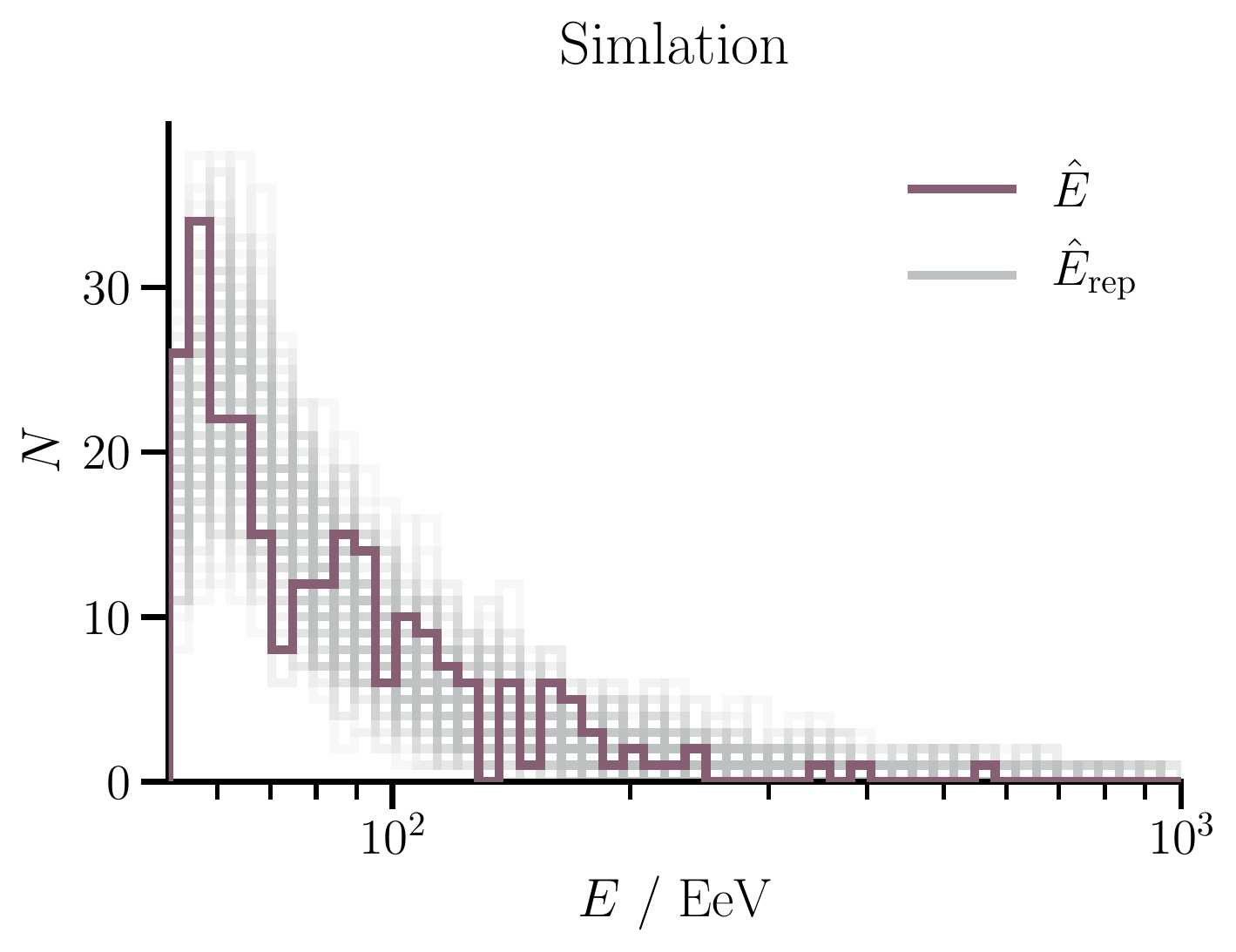}
\includegraphics[width=0.95\columnwidth]{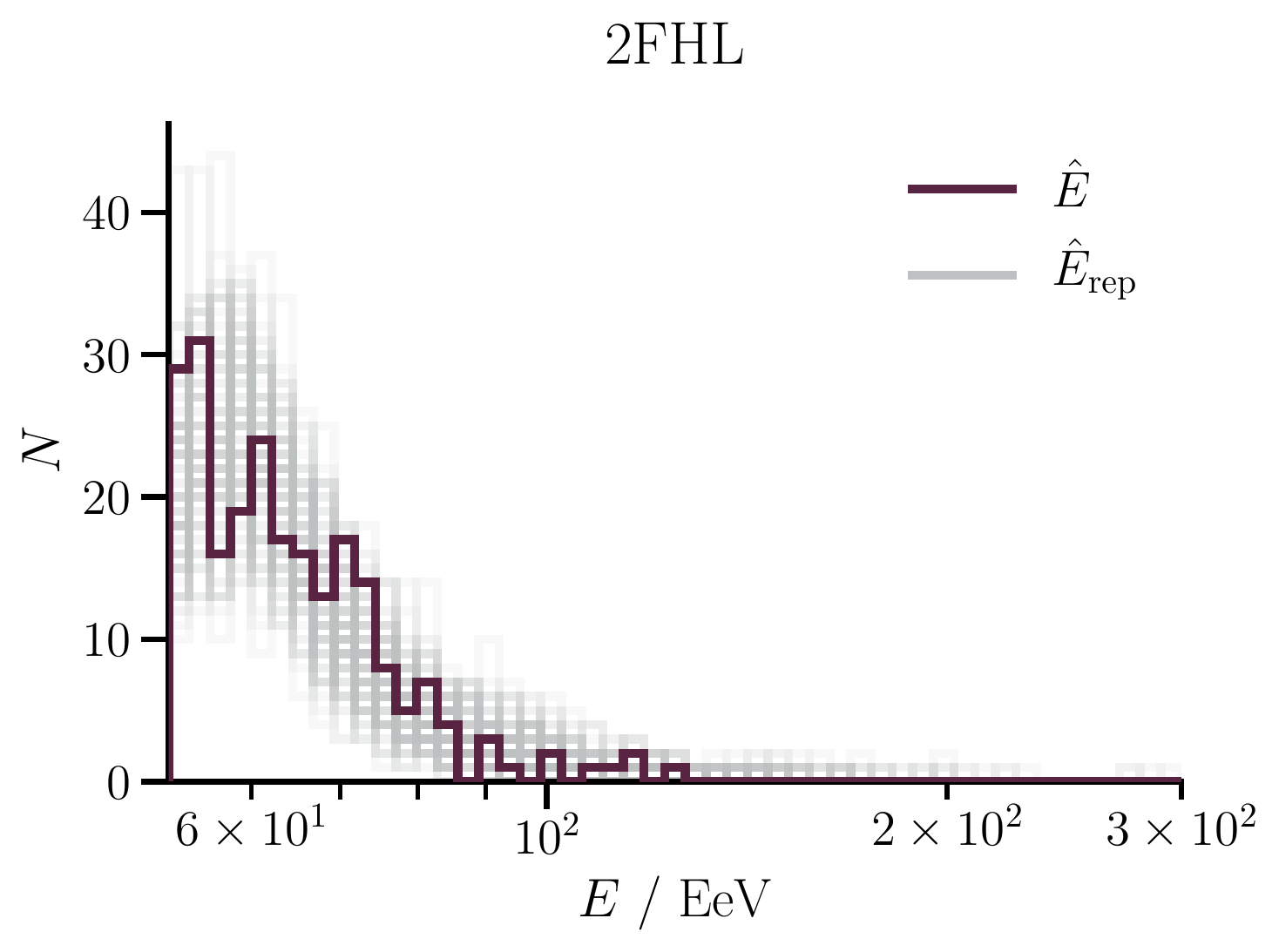}
\includegraphics[width=0.95\columnwidth]{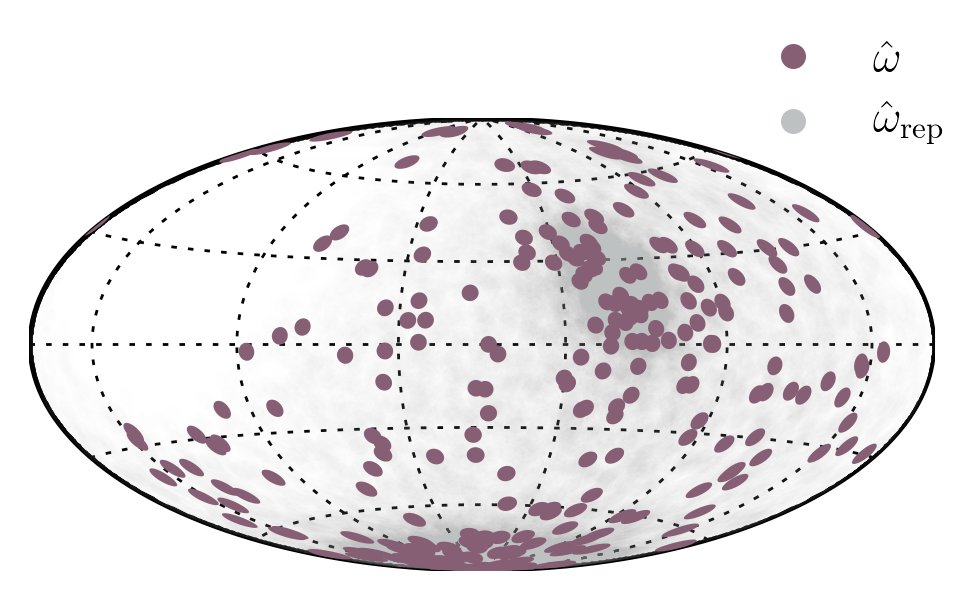}
\includegraphics[width=0.95\columnwidth]{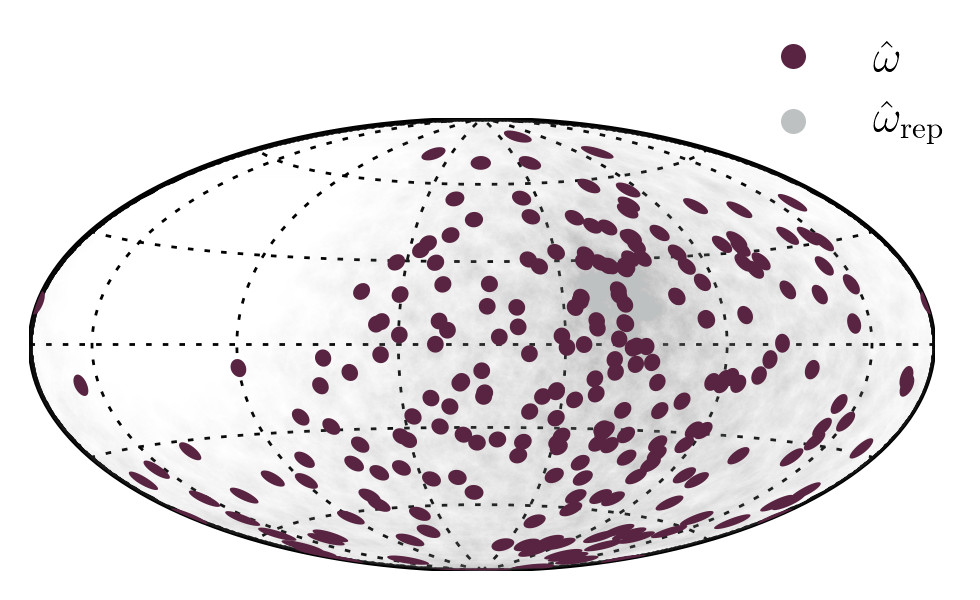}
\includegraphics[width=0.95\columnwidth]{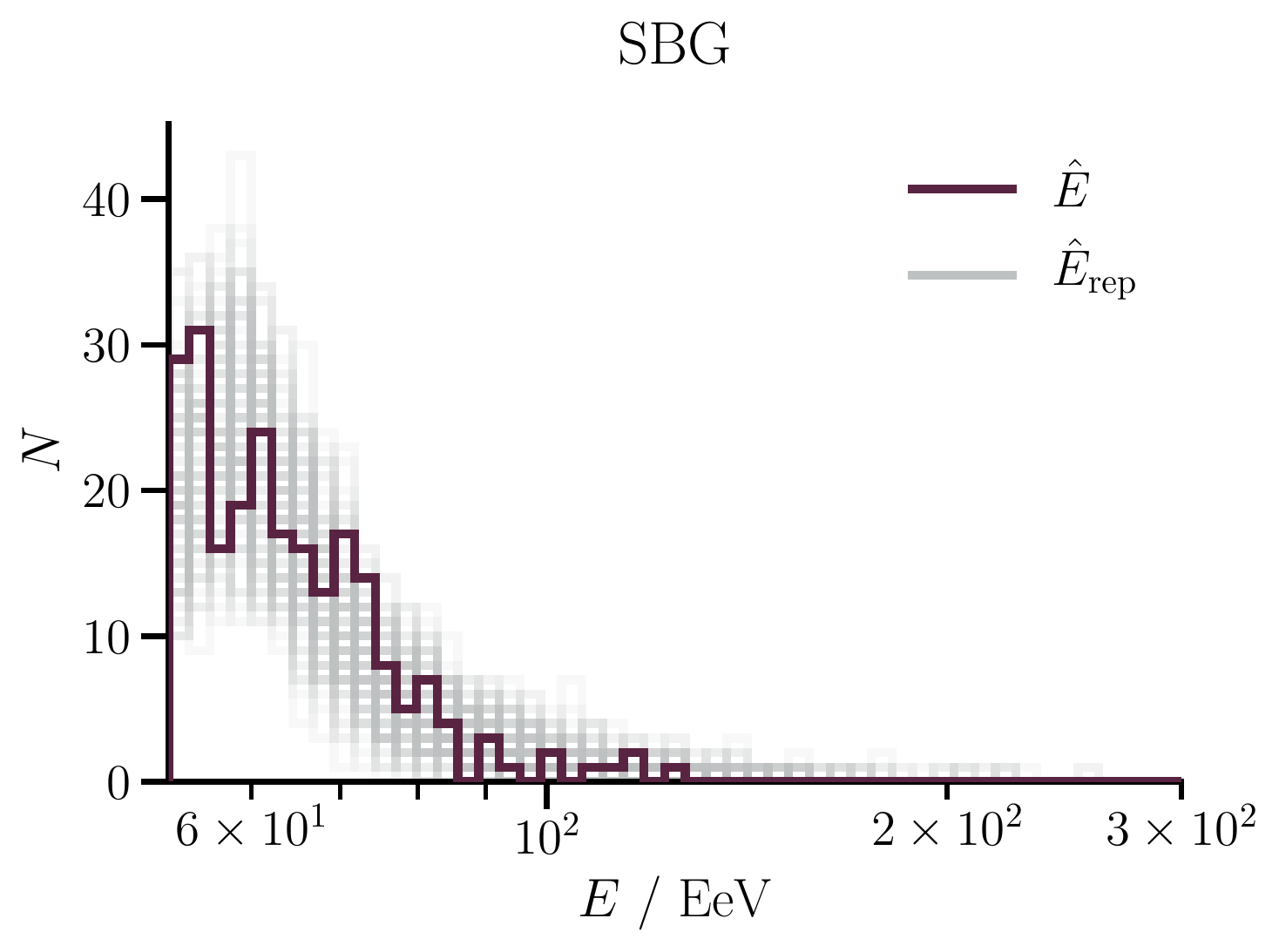}
\includegraphics[width=0.95\columnwidth]{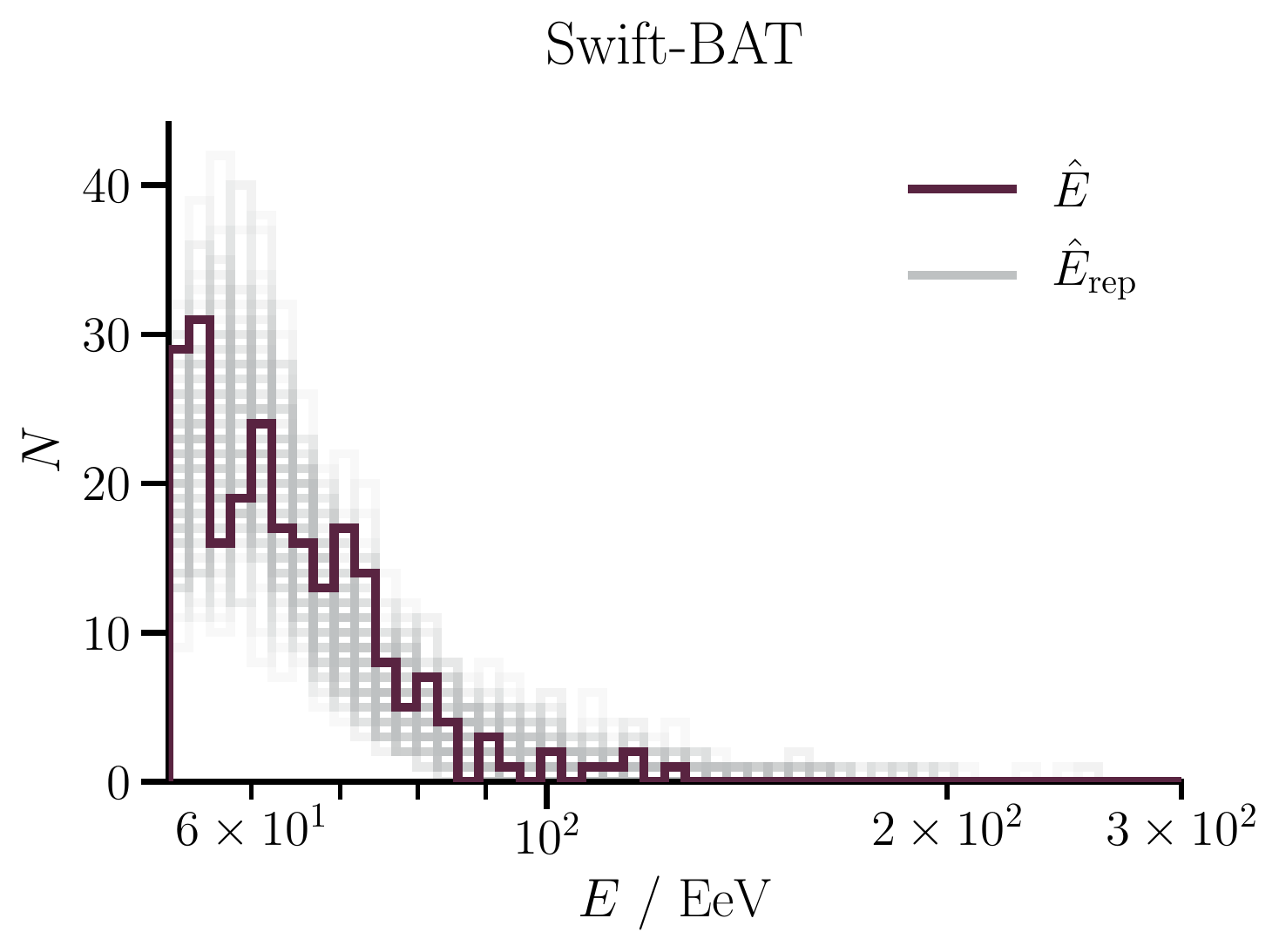}
\includegraphics[width=0.95\columnwidth]{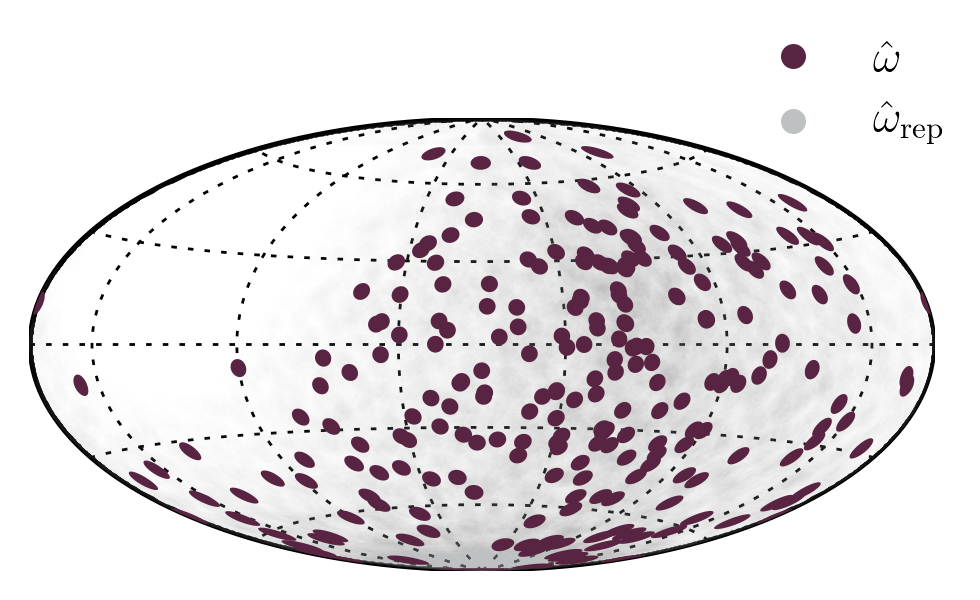}
\includegraphics[width=0.95\columnwidth]{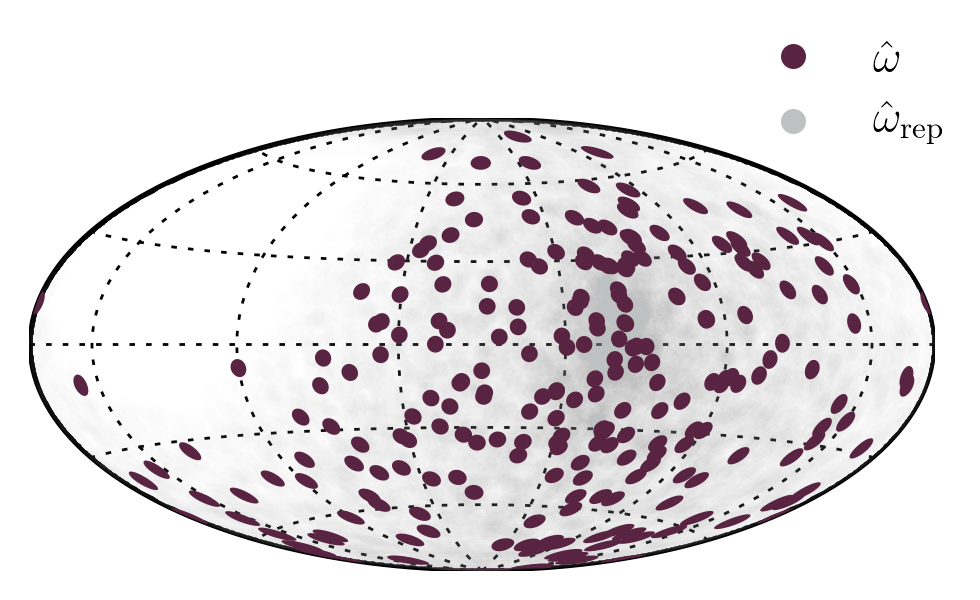}

 \caption{100 draws from the posterior predictive distribution, with the data, for the case of the simulated dataset (upper left) and fits to the PAO data using the 2FHL (upper right), SBG (lower left) and \emph{Swift}-BAT (lower right) catalogues. In each case, the energy spectrum and arrival direction distributions are shown separately with the data in solid colour and the replicated data overplotted in transparent grey.}
 \label{fig:ppc}
\end{figure*}

We compute 100 posterior predictive draws for both the fit of the simulated data discussed in Section \ref{sec:sim} and the fits of the PAO dataset with different source catalogues presented in Section \ref{sec:application}. The results are shown in Figure \ref{fig:ppc}. The simulated data shows a good fit, as is expected given that the model is the correct data generating process in this case. There are no systematic deviations between $\hat{E}$ and $\hat{E}_\mathrm{rep}$, and the clustering in the distribution of $\hat{\omega}$ is reflected in $\hat{\omega}_\mathrm{rep}$. For the fits to data, the $\hat{E}_\mathrm{rep}$ capture the overall shape of the detected energy spectrum but not the detail, and the slight shoulder in lower energies of $\hat{E}$ is missed. This could be due to modelling the UHECR energy spectrum with a simple power law with no cutoff and using the same spectral index for the source and background component, both of which are approximations used in this more simplistic model. The $\hat{\omega}_\mathrm{rep}$ distributions are more isotropic than that of the simulation, due to smaller association fractions being found. However, there is some slight clustering around the nearest source locations, as would be expected from the equal luminosity assumption.

These results are consistent with an associated fraction $f < 1$ due to some UHECRs coming from an unresolved background component due to distant sources, and others from sources which are not present in the chosen catalogues.

\bsp	
\label{lastpage}
\end{document}